\documentclass{aa}  
\usepackage{float}

\usepackage{graphicx}
\usepackage{txfonts}
\usepackage{ulem}
\begin{document}

   \title{Penumbral decay observed in active region NOAA 12585}

   \author{M. Murabito\inst{\ref{inst1}}, S.~L. Guglielmino\inst{\ref{inst2}},
   I. Ermolli\inst{\ref{inst1}}, P. Romano\inst{\ref{inst2}}, S. Jafarzadeh\inst{{\ref{inst3}, \ref{inst4}}}, L.~H.~M. Rouppe van der Voort\inst{{\ref{inst3}, \ref{inst4}}}}

   \institute{INAF -- Osservatorio Astronomico di Roma, Via Frascati,33 Monte Porzio Catone, RM, 00078, Italy\\
              \email{mariarita.murabitoi@inaf.it}\label{inst1}
              \and
              INAF -- Osservatorio Astrofisico di Catania, Via S. Sofia 78, I-95123 Catania, Italy\label{inst2}
              \and
              Rosseland Centre for Solar Physics, University of Oslo, P.O. Box 1029 Blindern, NO-0315 Oslo, Norway\label{inst3} 
              \and
              Institute of Theoretical Astrophysics, University of Oslo, P.O. Box 1029 Blindern, NO-0315 Oslo, Norway\label{inst4}}

   \date{}

% \abstract{}{}{}{}{} 
% 5 {} token are mandatory
 
  \abstract
   {The physical conditions leading the sunspot penumbra decay are poorly understood so far.   } 
   {We investigate the photospheric magnetic and velocity properties of a sunspot penumbra during the decay phase to advance the current knowledge of the conditions leading to this process.}
   {A penumbral decay was observed with the CRISP instrument at the Swedish 1m Solar Telescope on 2016 September~4 and~5 in active region NOAA 12585. During these days, full-Stokes spectropolarimetric scans along the \ion{Fe}{I} 630~nm line pair were acquired over more than one hour. We inverted these observations with the VFISV code in order to obtain the evolution of the magnetic and velocity properties. We complement the study with data from instruments onboard the Solar Dynamics Observatory and Hinode space missions.  }
   {The studied penumbra disappears progressively in both time and space. The magnetic flux evolution seems to be linked to the presence of Moving Magnetic Features (MMFs). Decreasing Stokes $V$ signals are observed. Evershed flows and horizontal fields were detected even after the disappearance of the penumbral sector. %counter-Evershed flows are found at some locations.
   }
   {The analyzed penumbral decay seems to result from the interaction between opposite polarity fields in type III MMFs and penumbra, while the presence of overlying canopies rules the evolution in the different penumbral sectors. }

   \keywords{Sun: photosphere - sunspots - Sun: magnetic fields -
             }
               
   \titlerunning{Penumbral decay observed in active region NOAA 12585}
   \authorrunning{M. Murabito et al.}
   
   \maketitle
%
%-------------------------------------------------------------------.

\section{Introduction}

The penumbra is an important part of a sunspot \citep[for a recent review see, e.g.][]{Tiwari17}, where the coupling between magnetic field and plasma is characterized by peculiar physical conditions due to magneto-convection. Therefore, it represents a fascinating and interesting feature. Although it has been the focus of many observational and theoretical studies, its nature is still not fully understood. This is the reason why penumbrae are among the science cases in the science critical plans for the next generation large-aperture solar telescopes, such as those for the Daniel K.~Inouye Telescope (DKIST, \citealp{RastDKIST}) and European Solar Telescope (EST, \citealp{Schlichenmaier19}). 

Observational and theoretical studies pointed out that the penumbra is characterized by radially aligned filaments hosting strong and weak magnetic fields interlaced with each other. The formation and decay processes of the penumbra occur when the vertical and horizontal magnetic field components change as reported in \citet{Balthasar13,Murabito2016} and \citet{Verma18}. In particular, it seems that the penumbra forms when the magnetic field reaches critical values for its inclination angle and strength (e.g., $\gamma \geq 60^{\circ}$ and $\mathrm{B}_{crit} \leq 1.6$~kG, \citealp{Rezai12}) and the magnetic flux exceeds $1-1.5 \times 10^{20}$~Mx \citep{Leka98,Rezai12}. Two competing scenarios have been proposed for penumbral formation: a bottom-up process, according to which emerging horizontal field lines become trapped by the overlying magnetic field and form the penumbra \citep{Leka98}, and a top-bottom process, where the magnetic fields, already emerged and forming a magnetic canopy above the proto-spot, change their inclination bending down to the photosphere with the onset of magneto-convection \citep{Romano14,Murabito2016}.

The physical conditions leading the penumbral decay are even more speculative. From both observational and theoretical points of view, three processes are believed to be involved during the decay phase: (i) displacement of \textit{Moving magnetic features} (MMFs), (ii) change of the inclination of the \textit{magnetic} field of the \textit{canopy} and (iii) \textit{convective motions}.

MMFs \citep{HarveyII73} are small magnetic concentrations mostly appearing around the penumbral outer boundary and moving radially outward in the moat region. They are classified into two groups: bipolar pair of MMFs (type I) or unipolar with the same/opposite polarity with respect to the parent spot. The unipolar ones are in turn divided into type II and III MMFs, if they are formed as a result of the detached penumbral spines, or as the intersections of the submerged penumbral flux, with same and opposite polarity with respect to the parent spot, respectively. \citet{Kubo2003} proposed that MMFs play a significant role in the penumbral decay. However, the limited spatial resolution of the observations reported in \citet{Kubo2003} did not allow those authors to understand whether type II or type III MMFs are responsible for the disintegration of the region. They only could speculate that, due to the larger number of type II MMFs detected, the latter could be the responsible for the removal of the magnetic field from the penumbra.
%Another interesting aspect regards the relationship between the MMFs and the flux removal from the sunspot. 
%Furthermore, the Bipolar MMFs indicate a serpentine field with multiple intersections with the solar photosphere. %Finally, the convection in the outer penumbra seems to be also related to the flux removal of the sunspot.

The change of inclination of the \textit{magnetic canopy} was invoked by \citet{Shimizu12} to explain the formation of the annular zone found in the chromosphere around a pore, before the appearance of the penumbra at photospheric level. \citet{Romano13,Romano14} confirmed this hypothesis, suggesting that the penumbra forms when the magnetic field lines of the magnetic canopy overlying the pore change their inclination, bending down to the photosphere. Conversely, it has been proposed that the straightening up of field lines belonging to the overlying magnetic canopy that disconnect from the photosphere can be conducive to the decay of sunspot penumbra \citep{Romano20}. Nevertheless, this process has not been observed yet.

%Shimizu et al. (2012) and Romano et al. (2013, 2014) observed at the chromospheric level the formation of an annular zone around the corresponding location of a pore in the photosphere, before the appearance of its penumbra. 
%Actually, the presence of canopy fields with a more horizontal configuration with respect to the solar photosphere has been proposed to lead to the formation of penumbral-like structures as well (e.g., Zuccarello et al. 2014; Guglielmino et al. 2017, 2019).

Numerical studies of the processes at work are also difficult. In general, it is a challenge to reproduce the decay process of a sunspot penumbra because it requires both the minimum resolution useful to resolve the penumbra details and a large simulation domain to capture the moat region where the sunspot is embedded. Despite this, some numerical models suggested a further mechanism for penumbral decay. Indeed, \citet{Rempel2015} simulated sunspots and naked spots (i.e., umbrae without surrounding penumbrae) to study how the presence of the penumbra affects the moat flow and the sunspot decay itself. He found that the sunspot decay may result from \textit{convective motions} placed deeper in the photosphere that erode the ``footpoint'' of the spot, probably leading to flux separation splitting the spot. Moreover, in the simulations, the presence of enhanced mixed polarity field in the moat region is found, whose strength depends on the strength of the overlaying magnetic canopy. Although the mixed polarity field arises from a different process than penumbral decay, it can contribute to the latter. In fact, a small imbalance between the submergence of horizontal (radial) magnetic field, as well as radial outward transport of vertical flux elements with same polarity as the spot, can be responsible for flux decay exceeding the standard sunspot flux decay rate of $10^{21} \,\mathrm{Mx \, day}^{-1}$ by an order of magnitude and leading to penumbral decay. In this respect, \citet{Kubo2008a} found that the \textit{G}-band bright features visible and moving in the outer penumbra, corresponding to MMFs separating from the penumbral spines, indicate the presence of subsurface upwelling and diverging flows that can destabilize the spot. This provides a possible link between decay owing to MMFs and erosion by convective motions.

In this study, we present unique high-resolution spectropolarimetric data of a decaying penumbra embedded in the bigger and complex magnetic flux system of active region (AR) NOAA 12585. Our observation of the penumbral decay adds a valuable piece of information to the rare evidence of this process, which allows us to analyse the role of the three above mentioned processes. 
Section~2 reports the data and their analysis, Section~3 presents the results and finally in Section~4 we provide our conclusion.

%--------------------------------------------------------------------
\section{Observation and data processing}

We used data acquired with different instruments at space- and ground-based telescopes, namely the Solar Dynamic Observatory \citep[SDO,][]{Pesnell2012} and the \textit{Hinode} \citep{Kosugi2007} satellites, as well as the Swedish 1-m Solar Telescope \citep[SST,][]{SST2003} in Canary Islands. The observations concern AR NOAA~12585, appeared at the East solar limb on 2016, September~1. In Table~\ref{table1}, the data used for our analysis are summarised. In the following, we will detail information about the space and ground-based, high-resolution datasets considered in our study.	

\begin{table*}[]
\center
\begin{tabular}{l l l l l l}
\hline 
\hline
                   &                     &                          &                         &               &  \\

\textbf{Telescope} & \textbf{Instrument} & \textbf{Spectral coverage}   & \textbf{Time coverage}   & \textbf{Spatial sampling} & \textbf{$\mu$}     \\ 
                   &               &                             &         & &   \\
\hline
     &                  &                                      &                 & &     \\
SST         &      CRISP       &  Fe I 630.2 nm line pair         & 2016/09/04  & 0\farcs06  &0.92   \\
  &     &             &         125 scans            & &    \\
 &    &               &    09:31:22-10:37:58 UT      & &   \\
                   &                     &                                                    &                              & &     \\
                   &                     &                                                    & 2016/09/05   &  & 0.98  \\
                   &                     &                                                    &          98 scans              &   &   \\
                   &                     &                                                    &       08:52:59-09:45:05 UT     &   &   \\
                   &                     &                                                    &                               &    & \\
SDO                &    HMI              &     Fe I 617.3 nm line    &  2016/09/03        & 0\farcs5  & 0.82  \\ 
                   &                     &                       &  2016/09/04       &            & 0.92   \\
                   &                     &                  &                   2016/09/05            &   &0.98 \\
                   &                     &                                                    &                               &      &    \\
HINODE             &     SP              &   Fe I 630.2 nm line pair                            & 2016/09/04                      & 0\farcs32  &0.92         \\
                   &                     &                                                    &   03:31-03:52 UT              &  & \\
                   &                     &                                                    &                              &   & \\
                   &                     &                                                    &2016/09/05        &  &0.98  \\
                   &                     &                                                    &  04:03-04:23 UT      &    &  \\
                   &                     &                                                    &                &      & \\
                   &                     &                                                    & 2016/09/05           &  & 0.98\\
                   &                     &                                                    &   07:17-07:37 UT             &   & \\
                   &                     &                                                    &                              & & \\
%                   &                     &                                                    & 05 - 09 @ %14:14-14:46 UT       &                  \\
\hline
\end{tabular}
\label{TAB1}
\caption{Ground- and space-based observations exploited in this study.   %\textbf{i dati delle 14 sono stati rimossi dall'analisi perche la penombra è già bella che andata!}
}\label{table1}
\end{table*} 

%\textcolor{blue}{\textbf{Rimuovere dalla caption della tabella il fatto che i dati di SST sono in BOLD; Togliere anche POL dalla tabella}}

\subsection{SDO/HMI and Hinode/SP datasets}

Data taken by the Helioseismic and Magnetic Imager \citep[HMI,][]{Scherrer2012} on board the SDO satellite were used. In particular, we analyzed Space weather HMI AR Patches \citep[SHARPs,][]{Bobra2014} continuum filtergrams and maps of the radial component of the vector magnetic field (B$_{r}$). These data were acquired along the \ion{Fe}{I} 617.3~nm line with a spatial resolution of 1\arcsec, with a cadence of 12 minutes over five days of observations. We considered data relevant to AR NOAA~12585 taken from 2016 September~1 at 00:00~UT to September~5 at 23:48~UT.

The \textit{Hinode}/Solar Optical Telescope \citep[SOT,][]{Tsuneta2008} performed three raster scans over part of the AR NOAA~12585 with the spectropolarimeter \citep[SP,][]{Lites2013}, from 03:31~UT on September~4 to 07:37~UT on September~5. The duration of each scan was about 20 minutes. Full Stokes parameters were acquired along the \ion{Fe}{I} at 630.2~nm line pair, with a pixel size of 0\farcs32 (fast map mode) and a variable field of view (FOV). A subFOV centered in the region of interest (ROI) of $45\farcs3 \times 64\farcs3$ was selected (see the blue box in Fig.~\ref{fig:fig1}). We normalized these measurements by the average intensity in the continuum of the \ion{Fe}{I} 630.2~nm line pair in a large quiet-Sun region placed in the upper part of the \textit{Hinode}/SP FOV. 
We also corrected the perturbation introduced by the Point Spread Function (PSF) of the telescope by applying the regularization method proposed by \citet{Ruiz2013} and used in \citet{Quintero2015, Quintero2016,Guglielmino2018,Guglielmino2019}. This method is based on a principal component decomposition of the Stokes profiles.
In particular, we considered the first 8 principal components of eigenvectors for Stokes~I and~V and 4 for the remaining Q and~U profiles as in \citet{Quintero2015}, with 15~iterations, using the spatial PSF for SOT/SP measurements with 0\farcs32 pixel size. The continuum contrast in a large quiet-Sun region increases from 6.5\%, 6.2\% and 6.3\% for the three original maps, to 13.9\%, 12.8\% and 13\% in the deconvolved maps, respectively. These values are similar to those obtained in previous works \citep{Quintero2015,Quintero2016,Guglielmino2019}. 

We derived the Circular polarization (CP) and Linear polarization (LP) maps by adapting the routine for the level 1.5 \textit{Hinode}/SP data available in the Community Spectropolarimetric Analysis Center\footnote{http://www2.hao.ucar.edu/csac} to the deconvolved data. Finally, the line-of-sight (LOS) velocity field has been derived from the inversion of the \textit{Hinode}/SP deconvolved data using the Very Fast Inversion of the Stokes Vector code \citep[VFISV,][]{Borrero2011}, version 4.0, which performs a Milne-Eddington inversion of the observed Stokes profiles. Then, we calibrated the LOS velocity by imposing that the plasma in a quiet Sun region has on average a convective blueshift \citep{Dravins1981}. Following \citet{Balthasar1988}, the average convective blueshift velocity for the \ion{Fe}{I} 630.25~nm at the $\mu$ positions reported in Table~\ref{table1} is equal to $-97$ and $-126$ $\mathrm{m \, s^{-1}}$ for September~4 and~5, respectively.

%applying the Gaussian fits to the line profiles of the Fe I 630.15 nm line (using the first 8 spectral points). We reconstructed the profiles of this line in each spatial pixel by fitting the corresponding Stokes I component with a linear background and a Gaussian-shaped line. Finally, the LOS velocity values were deduced from the Doppler shift of the centroid of the line profiles in each spatial point in comparison to the nominal wavelength.} 

\subsection{SST dataset}

\begin{figure*}[t]
	    \includegraphics[scale=0.39,clip,trim=50 635 250 30]{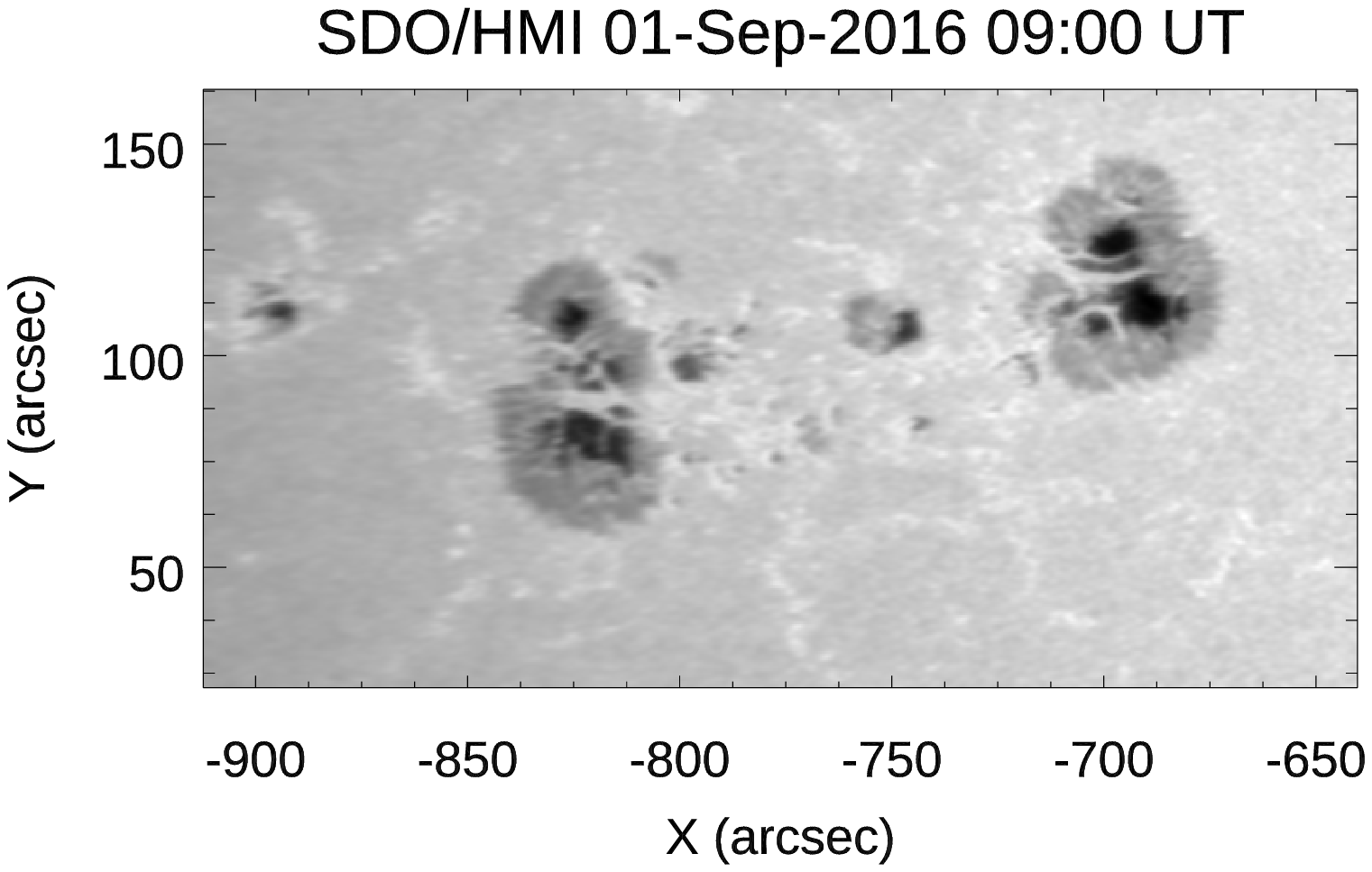}
	    \includegraphics[scale=0.39,clip,trim=50 635 250 30]{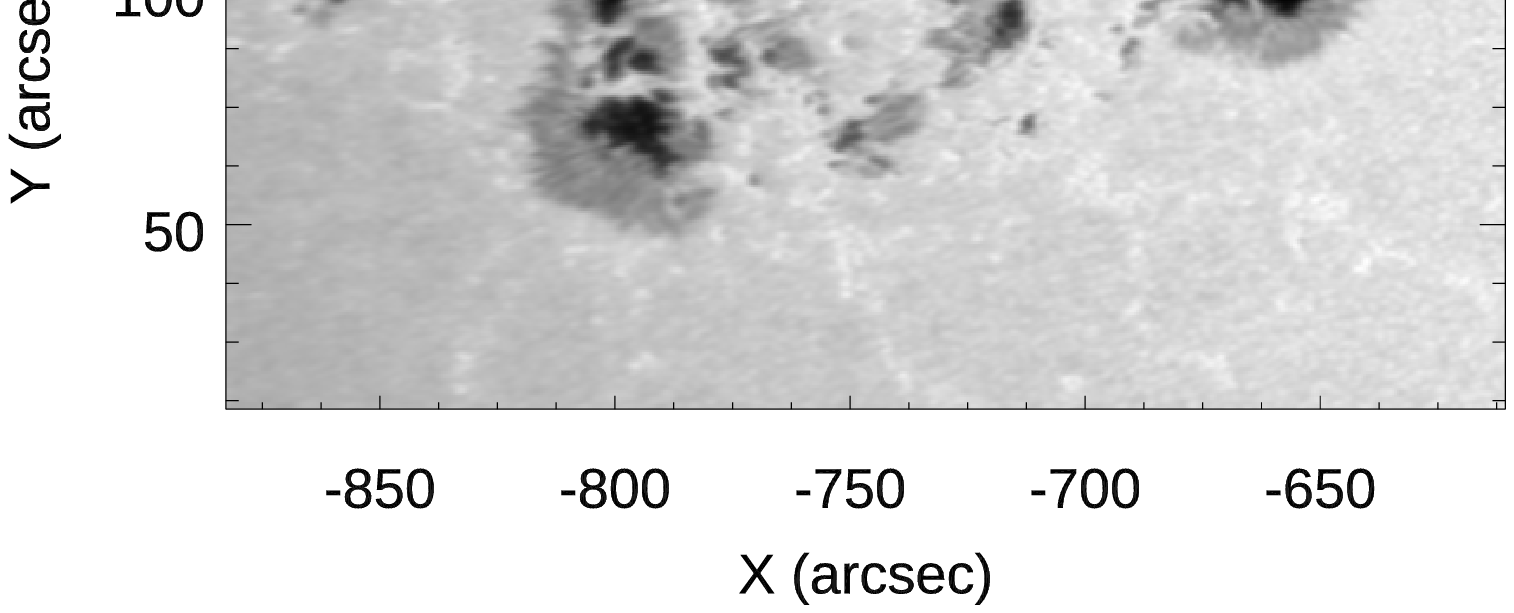} 
        \includegraphics[scale=0.39,clip,trim=50 635 250 30]{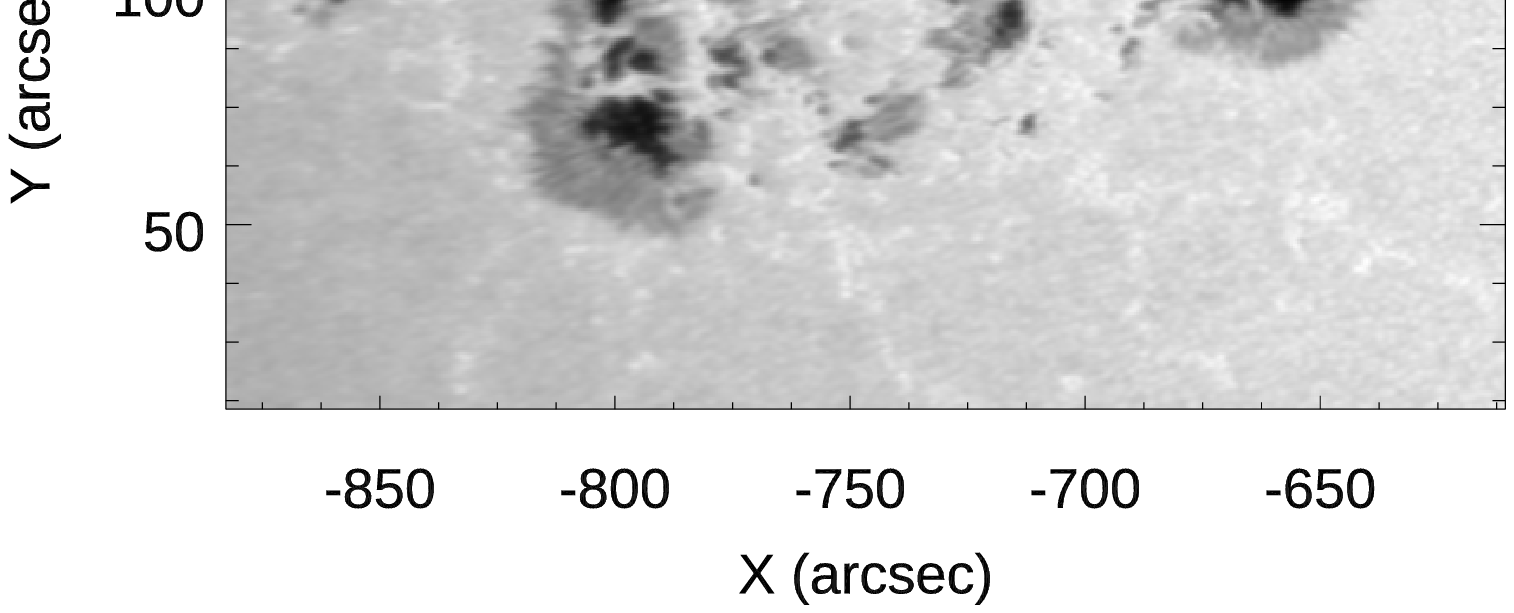}\\
	\includegraphics[scale=0.39,clip,trim=50 620 250 30]{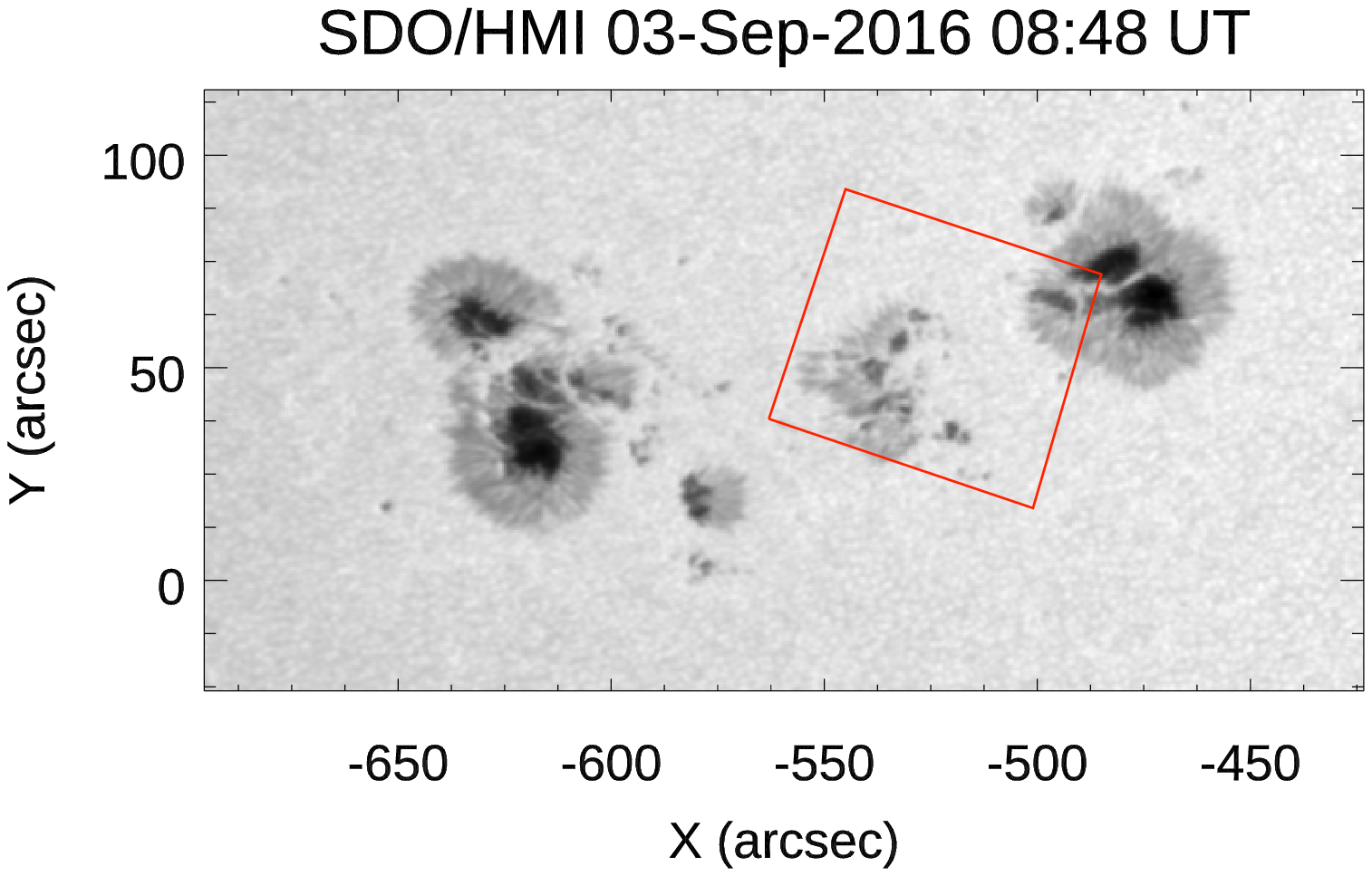}
	\includegraphics[scale=0.39,clip,trim=50 620 250 30]{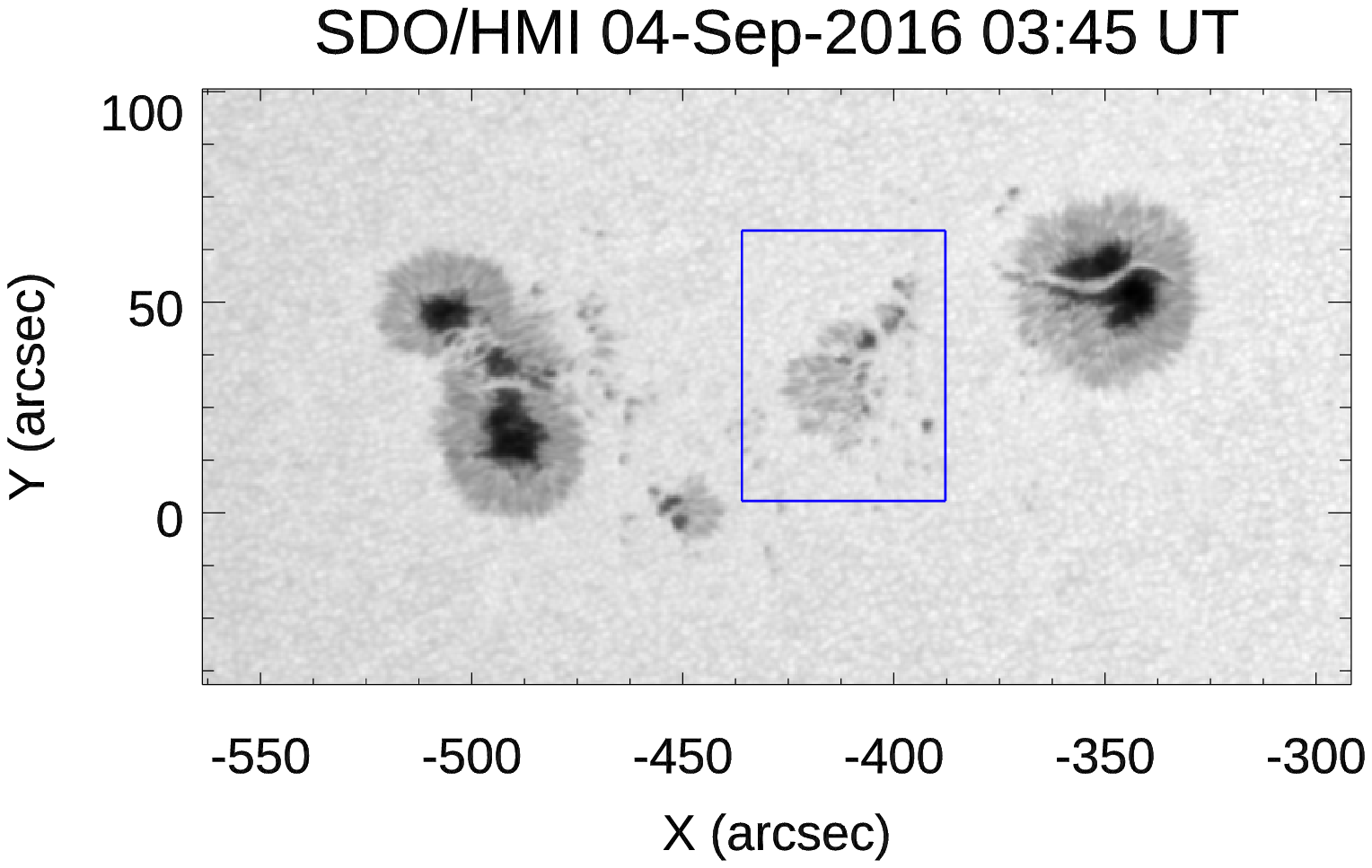} 
	\includegraphics[scale=0.39,clip,trim=50 620 250 30]{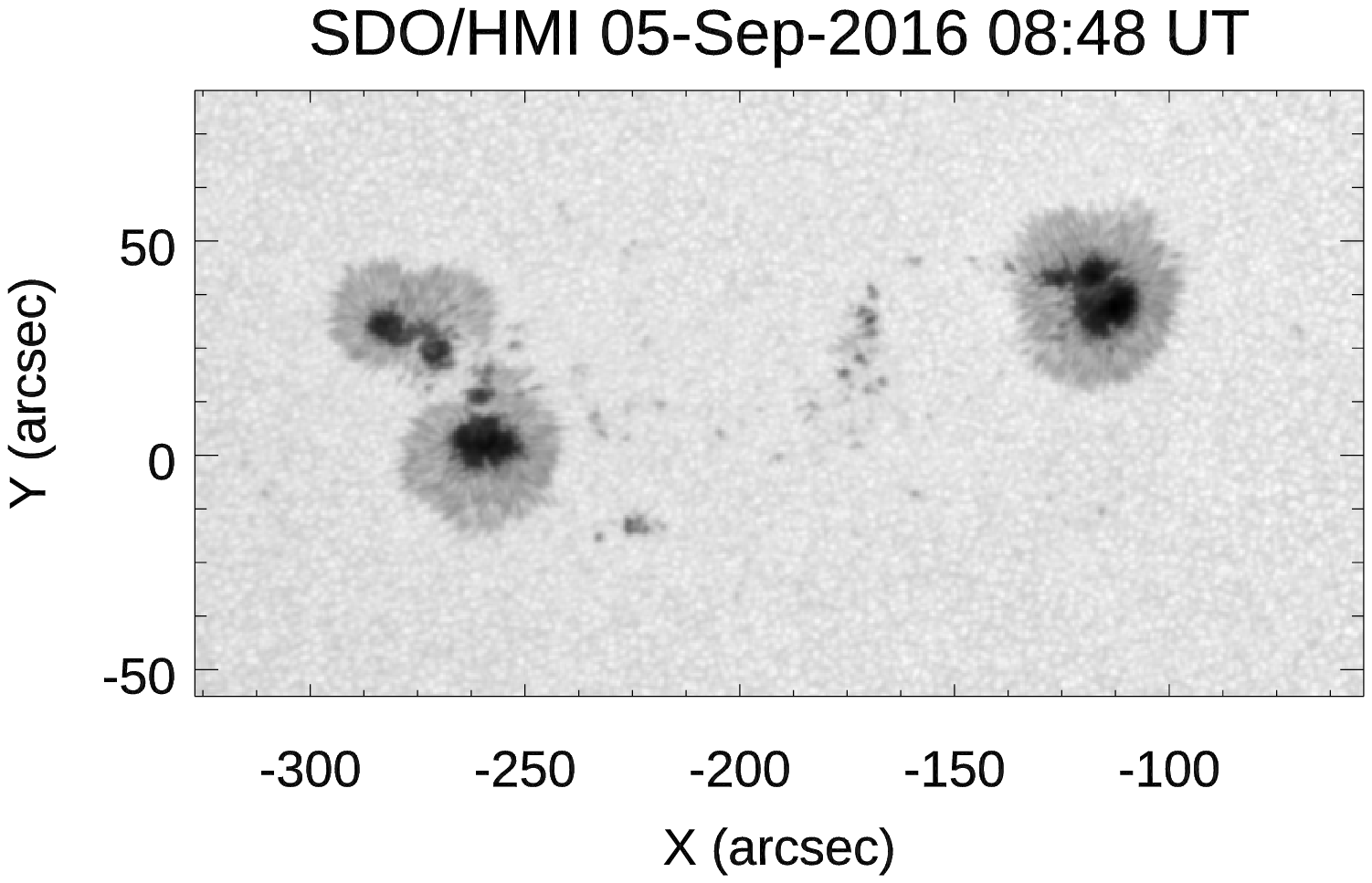}\\
	\caption{Maps of the continuum intensity of AR NOAA 12585 acquired by SDO/HMI. The red and blue boxes indicate the SST/CRISP and \textit{Hinode}/SP FOVs analyzed in this study, respectively.}
	\label{fig:fig1}
\end{figure*}

Part of AR NOAA~12585 was observed on September~4 and~5 with SST using the CRisp Imaging SpectroPolarimeter \citep[CRISP,][]{CRISP}. See the red box in Fig.~\ref{fig:fig1} for the Field of View (FOV) of the observations sampled by SST.  For the sake of clarity, the box is overplotted to the SDO/HMI subFoV from September 3 data.   

CRISP carried out spectropolarimetric measurements along the \ion{Fe}{I} 630.2~nm line pair, \ion{Ca}{II} at 854.2~nm and spectroimaging observations along the profile of the H${\alpha}$ line at 656.3~nm. The \ion{Fe}{I} 630.2~nm  line pair was sampled along 16 spectral positions between 630.13 and 630.26~nm, with a temporal cadence for each complete scan of $\sim 32$~s. The pixel scale of these measurements is 0\farcs06. In this study, we analyze the photospheric data, deferring the analysis of chromospheric observations to future work. 

The data were processed with the standard reduction pipeline \citep{delacruz2015}. This includes image restoration with the Multi-Object Multi-Frame Blind Deconvolution method \citep[MOMFBD,][]{vannoort2005}. These data have also been described and analysed by \citet{Rvandervoort17} and \citet{Ortiz2020}. We inverted the CRISP data with the VFISV code. Finally, we corrected the inclination for the $180^{\circ}$-azimuth ambiguity and the vector field components transformed into the local solar frame using the non-potential field calculation code \citep[NPFC,][]{Georgoulis2005}. 
%We estimated the LOS velocity by applying a Gaussian fit to the line profiles of the \ion{Fe}{I} 630.15~nm line, using the first 8 spectral points. In more detail, we reconstructed the profiles of this line in each spatial pixel by fitting the corresponding Stokes~I component with a linear background and a Gaussian-shaped curve. Finally, we deduced the LOS velocity values from the Doppler shift of the centroid of the line profiles in each spatial point in comparison to the nominal central wavelength. 
For the velocity calibration of SST/CRISP data we applied the same procedure used for the \textit{Hinode}/SP observations. In particular, taking into account the $\mu$ positions reported in the Table~\ref{table1}, we considered an average convective blueshift velocity for the Fe I 6301.5 equal to $-195$ and $-229$  $\mathrm{m \, s^{-1}}$ for September~4 and~5, respectively.

We computed the CP and LP signals, pixel-wise, following the method set out by \citet{MartinezPillet2011} and adapted to our specific dataset, as reported in \citet{Murabito17}. For our \ion{Fe}{I} 630.15~nm line data, we consider:
\begin{eqnarray}
CP=\frac{1}{7 \left \langle I_{c} \right \rangle} \sum_{i=1}^{7} \epsilon_ {i} {V_{i}}  \\
LP=\frac{1}{7 \left \langle I_{c} \right \rangle} \sum_{i=1}^{7} \sqrt{Q^{2}_{i}+U^{2}_{i}}
\end{eqnarray} 

\noindent
where $\epsilon$ = 1 for the first three spectral positions of the line sampling (i.e. on the blue wing), $\epsilon$ = -1 for the last three positions (i.e. on the red wing), and $\epsilon$ = 0 for the line center position, while \textit{i} runs from the 1st to the 7th wavelength position.

\begin{figure*}[t]
	    \includegraphics[scale=0.39,clip,trim=50 635 250 30]{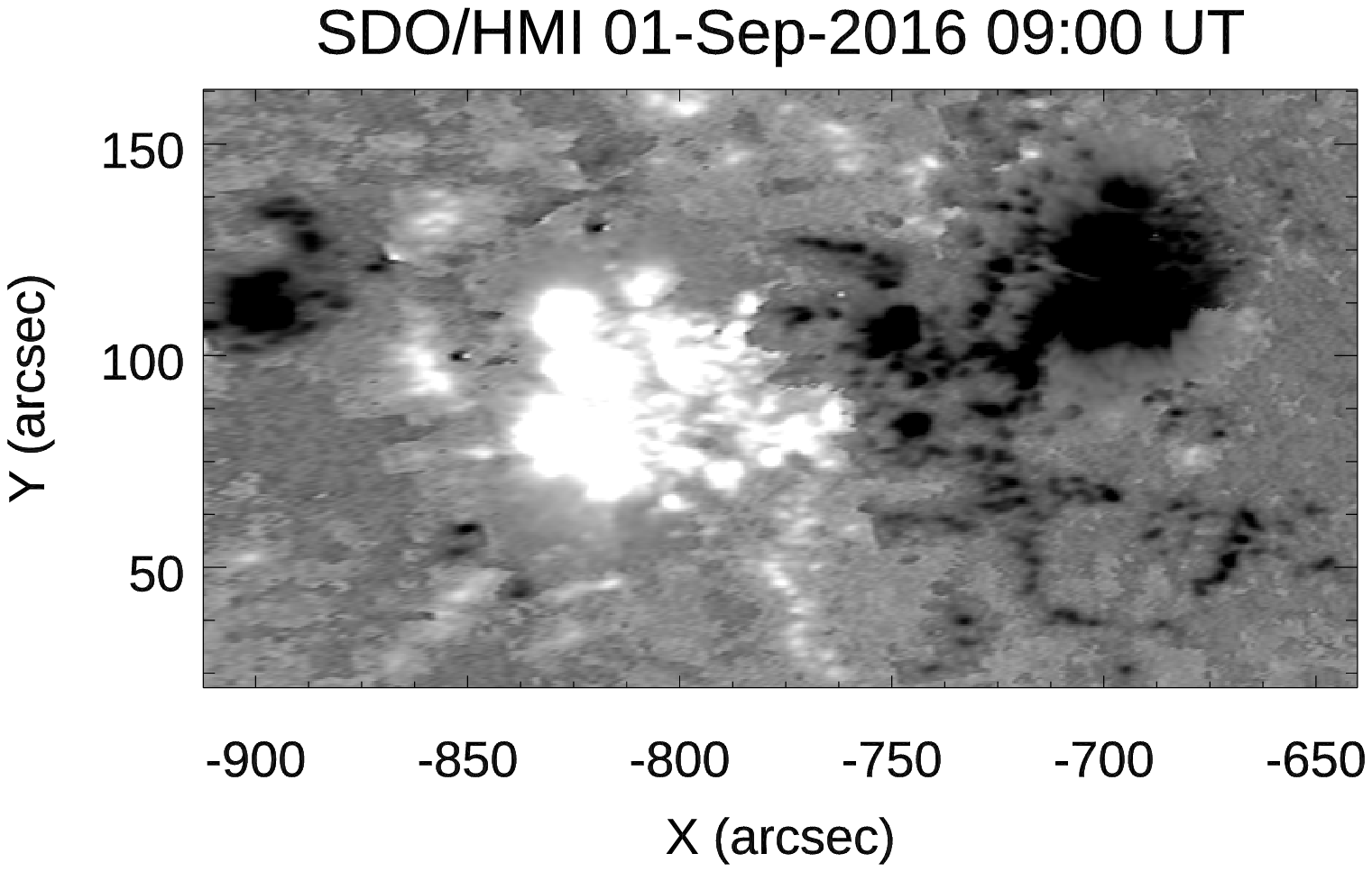}
	    \includegraphics[scale=0.39,clip,trim=50 635 250 30]{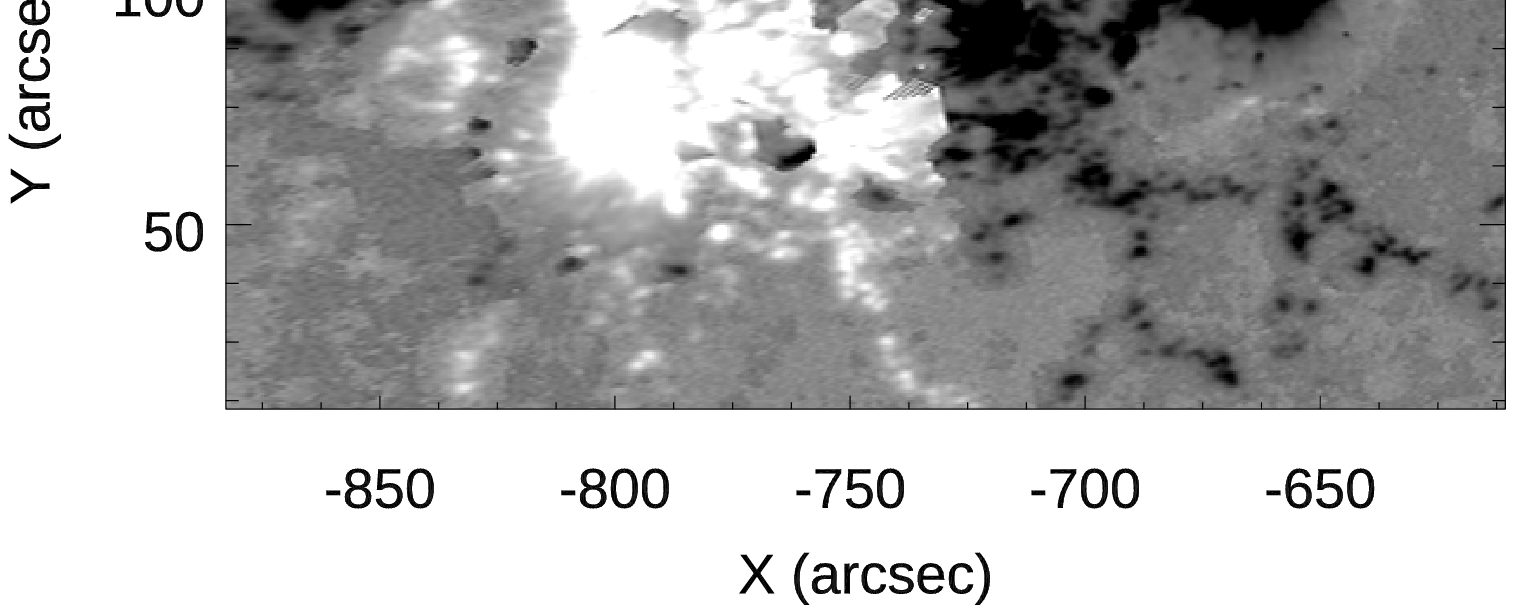} 
	    \includegraphics[scale=0.39,clip,trim=50 635 250 30]{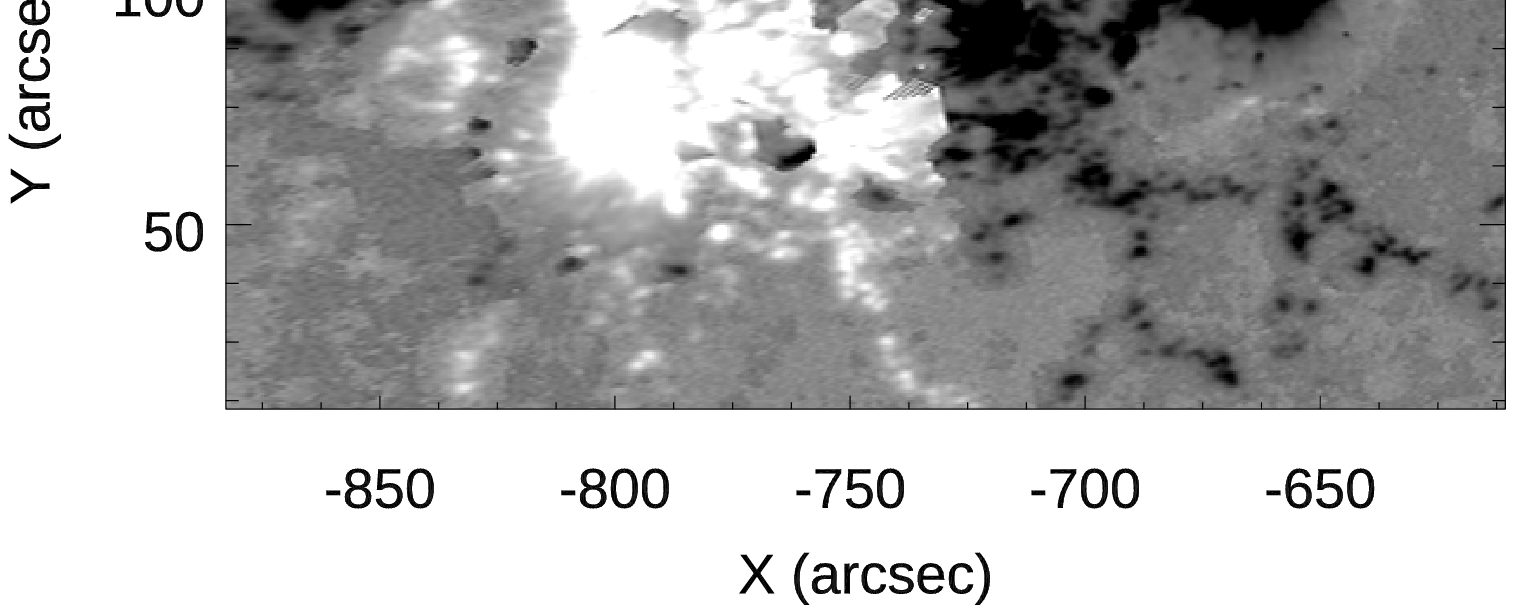}\\
	\includegraphics[scale=0.39,clip,trim=50 620 250 30]{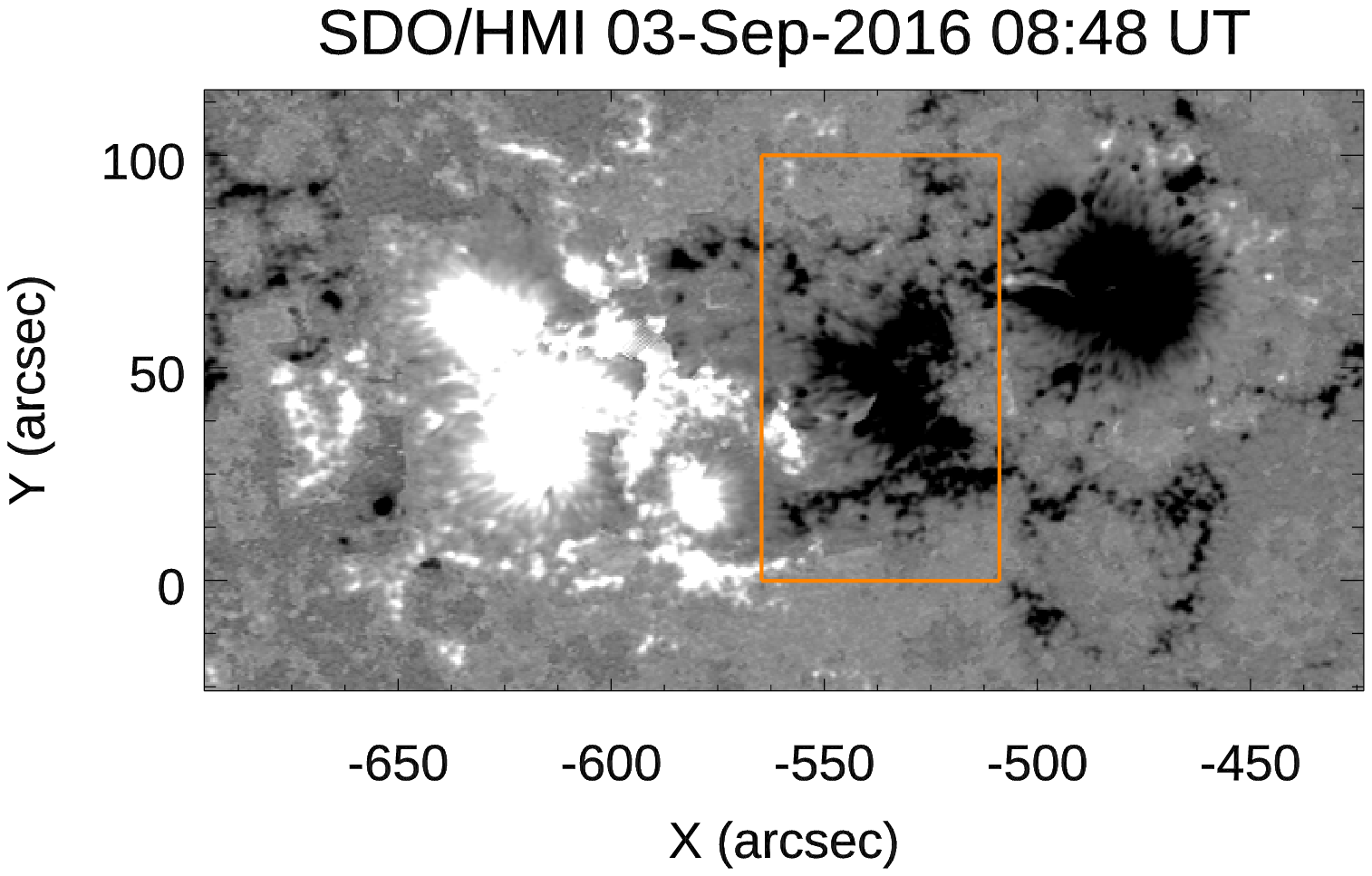}
	\includegraphics[scale=0.39,clip,trim=50 620 250 30]{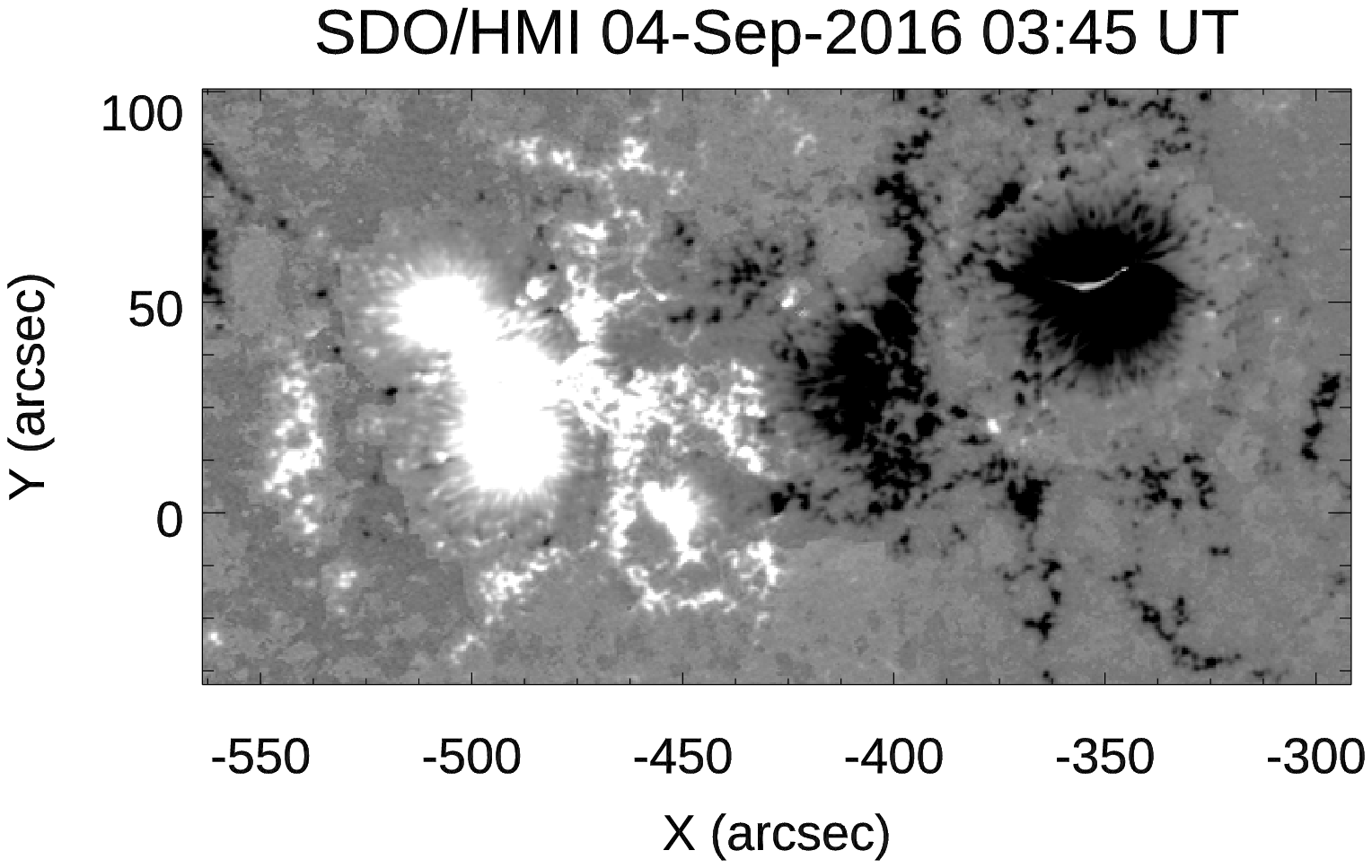} 
	\includegraphics[scale=0.39,clip,trim=50 620 250 30]{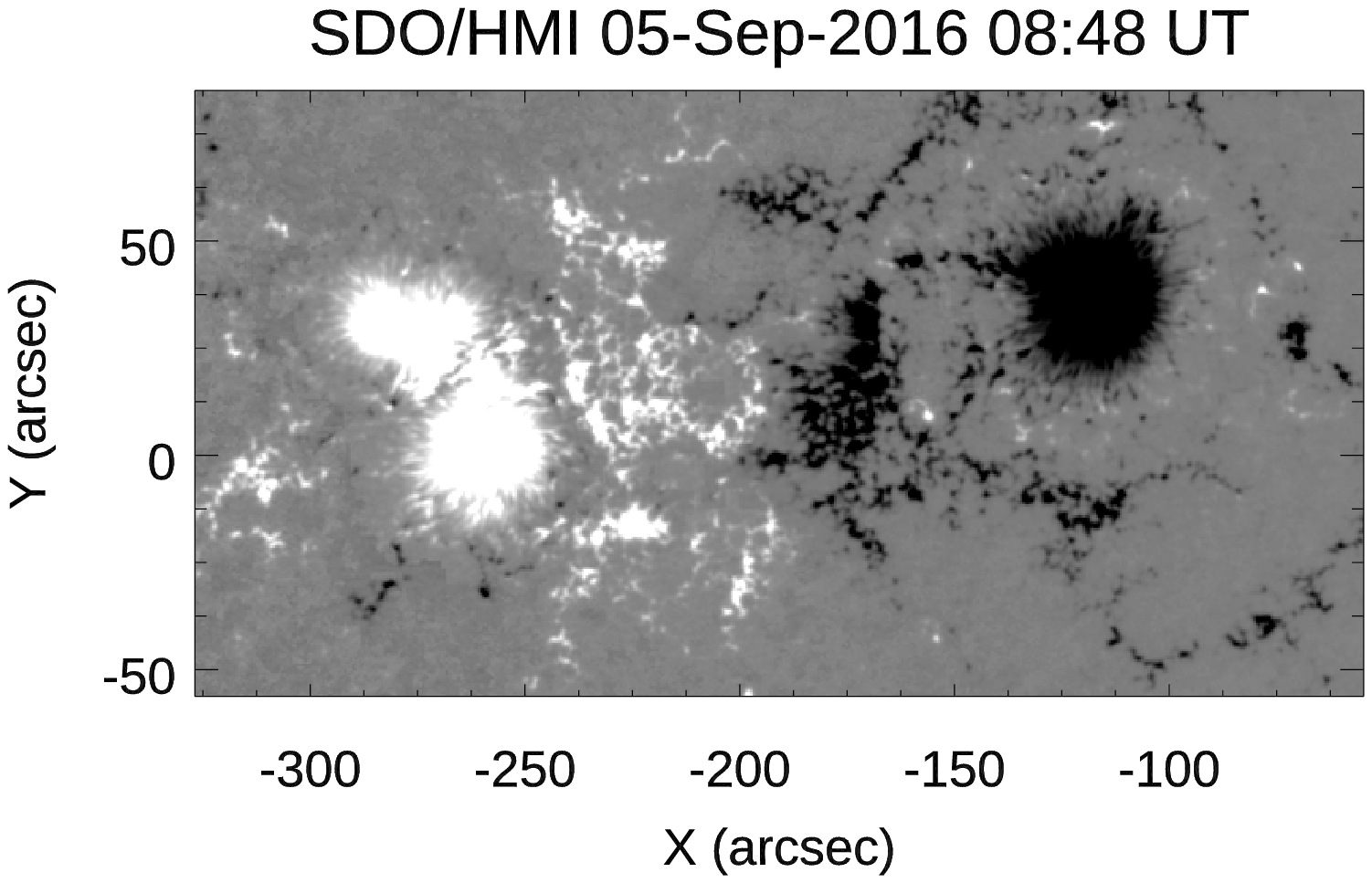}\\
	\caption{Maps of the LOS magnetograms of AR NOAA~12585 acquired by SDO/HMI. The orange box indicates the FOV where the magnetic flux evolution shown in Figure~\ref{fig:fig2} has been computed.}
	\label{fig:fig1bis}
\end{figure*}

\section{Results}

\subsection{Global evolution of AR NOAA 12585}

Figures~\ref{fig:fig1} and~\ref{fig:fig1bis} display the large-scale evolution of the AR from September~1 to~5 in the continuum filtegrams and radial magnetograms as acquired by SDO/HMI. The AR is initially characterized by a $\beta$ magnetic configuration and mainly composed by two complex sunspots (see top panels of Fig.~\ref{fig:fig1}). These latter exhibit light bridges (see, e.g., \citealp{Falco16,Felipe16} and references therein) during the observing time interval, indicating the onset of their fragmentation. It is worth mentioning that the light bridge of the leading sunspot has polarity opposite than the hosting umbra, as recently observed in umbral filaments \citep{Guglielmino2017,Guglielmino2019}. On September~5, the leading spot displays a smaller size but a more roundish shape with a regular penumbra, while the trailing polarity after a fragmentation process splits in two regions with several umbral cores embedded in more irregular penumbrae (see bottom panels of Fig.~\ref{fig:fig1}). 

The central region in between the two main AR polarities appears as composed by smaller and more fragmented magnetic structures (see the magnetograms of Fig.~\ref{fig:fig1bis}) showing the formation of two not fully developed sunspots. As a consequence, the magnetic configuration of the AR changes from $\beta$ to $\beta \gamma$ on September~3. From September~1 to~3, this portion of the AR evolves in several pores, forming through the reorganization of the emerged flux by the combined effect of velocity and magnetic field (see \citealp{Ermolli2017}), and a more extended penumbra. Interestingly, this penumbra forms only eastwards, i.e., towards the opposite polarity, while no penumbra forms on the other side. This sub-system starts to decay before it was able to form a complete penumbra. 

\begin{figure}
	\centering
	\includegraphics[scale=0.42,clip,trim=110 50 110  190]{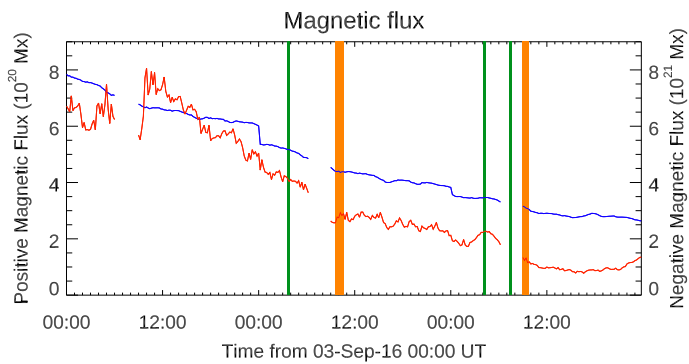}
	\caption{Evolution of positive (red) and negative (blue, absolute value) magnetic flux in the ROI (marked with the orange box in Fig.~\ref{fig:fig1bis}) as derived from SDO/HMI measurements. The orange stripes represent the time of SST/CRISP observations, while the green ones those of the \textit{Hinode}/SP scans.}
	\label{fig:fig2}
\end{figure}

\subsection{Study of the decay phase of the central region}

\begin{figure*}
    \includegraphics[scale=0.6,clip,trim=100 360 190  0]{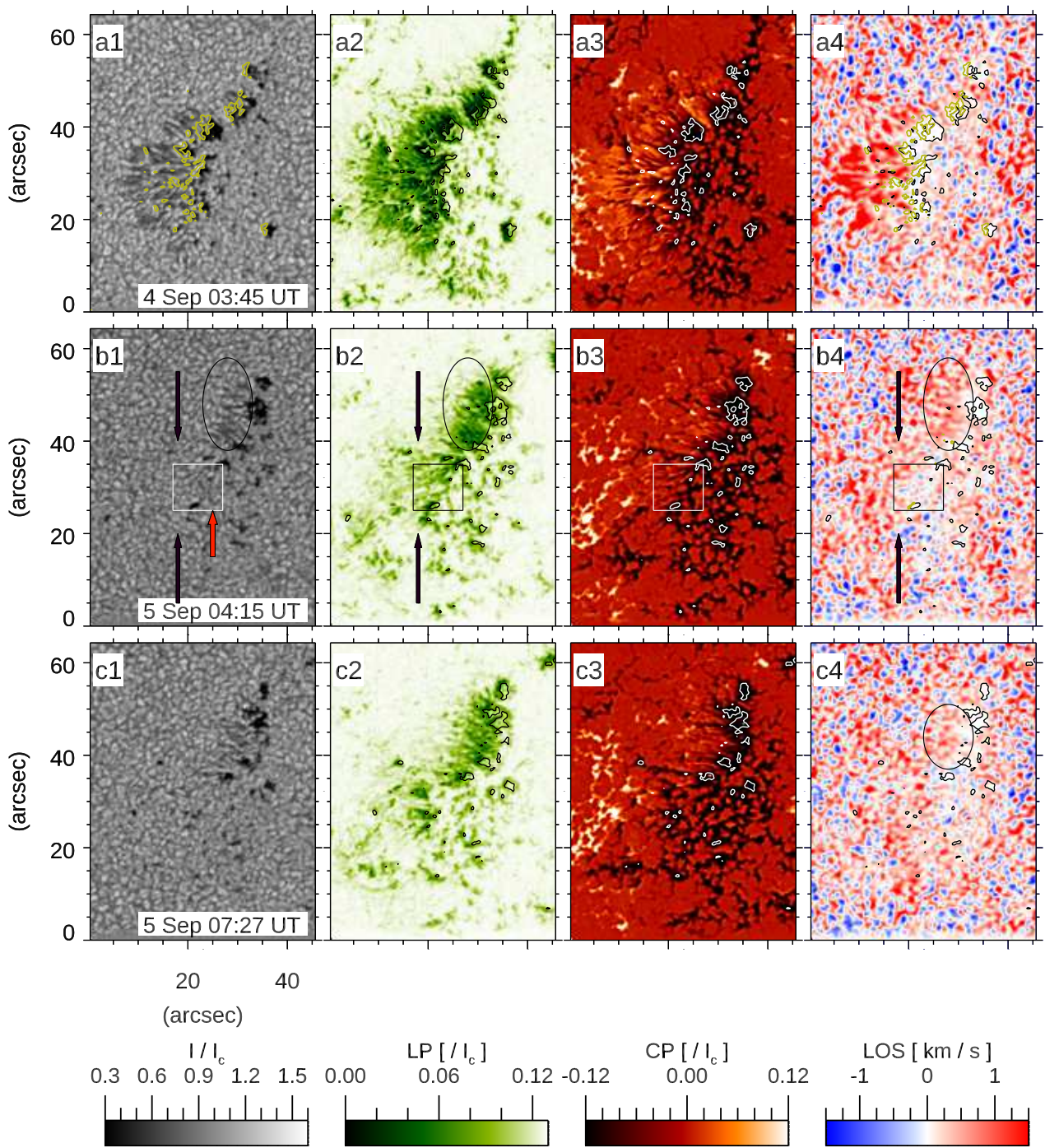}
    \caption{From left to right: Continuum intensity, linear polarization LP, circular polarization CP and LOS velocity relevant to the data acquired during the three \textit{Hinode}/SP raster scans. The subFOV is the same as the blue box shown in Fig.~\ref{fig:fig1}. The black/white contours represent the edge of the umbra ($I_{c}=0.6$) as seen in the continuum intensity. The light green contour in the first continuum and LOS velocity maps indicates contour at $0.1$ $LP / I_{c}$. Arrows, ovals and boxes point to regions and features described in Sect-~3.2. }
    \label{fig:fig_hinode1}
\end{figure*}

We computed the variation of the positive and negative magnetic flux  (from SDO/HMI) in the subFOV indicated by the orange box in Fig.~\ref{fig:fig1bis} (bottom-left panel), from September~3 and~5. The plot reported in Fig.~\ref{fig:fig2} clearly shows the decay phase of the central part of the AR. Figure~\ref{fig:fig2} also displays the time intervals covered by the SST/CRISP and \textit{Hinode} datasets, indicated with the orange and green stripes, respectively. To estimate the decay rate we calculated $d\Phi/dt$, where $\Phi$ represents the total flux corresponding to the sum of the positive and negative (in absolute value) fluxes during the entire time interval. From September~3 to~5, the flux in this region decayed with an average rate of $2 \times 10^{16} \,\mathrm{Mx \,s}^{-1}$. \\
Figure~\ref{fig:fig_hinode1} displays the reconstructed continuum intensity (panels a1, b1 and c1), the LP (panels a2, b2 and c2), CP (panels a3, b3 and c3), and LOS velocity (panels a4, b4 and c4) maps derived from the measurements taken by \textit{Hinode}/SP, respectively at three time steps on September~4 and~5. \\
The high spatial resolution of the \textit{Hinode} observations compared to that from SDO, allows seeing, on September~4, the fine structure of the penumbra and its discontinuities: e.g., penumbral grains and patchy bright features are visible at X,Y=[20\arcsec,20\arcsec] or [15\arcsec,30\arcsec]. These areas, which are characterized by a higher continuuum intensity, could be interpreted as the manifestation of the intrusion of granulation among the penumbral filaments. Furthermore, these areas display the lowest value of LP signal. %They seems to correspond to free-inclined magnetic fields as indicated by the LP map.}
Besides, the CP map at 03:45~UT displays positive and negative patches inside the filamentary penumbral structure of the field, resembling MMFs. \\
Comparing the \textit{Hinode}/SP maps at different times, we see that the LP maps show a decrease of the horizontal field with time. Moreover, the CP signal that outlines the fine structure of the penumbra becomes weaker as penumbral filaments disappear. The granulation seems fuzzy where the field starts to dissipate and the penumbra dissolves (see the red arrow in panel b1 of Fig.~\ref{fig:fig_hinode1}). In these regions, the field appears to be mainly horizontal on September~5 at~04:15 and~07:27~UT (see the panels a2, b2 and c2 of Fig.~\ref{fig:fig_hinode1}), while the continuum intensity maps show the re-organization of the granulation in the southern side of the ROI (see the panel b1 of Fig.~\ref{fig:fig_hinode1}).

The LOS velocity maps display that the penumbral region is initially redshifted according to the expected classical Evershed flow pattern (see panel a4 of Fig.~\ref{fig:fig_hinode1}), being located towards the eastern limb. The Evershed flow is visible \textbf{as long as} some penumbral filaments remain (see the ovals in the panels b4 and c4 of Fig.~\ref{fig:fig_hinode1}). The granulation velocity pattern in the velocity maps takes more time to re-appear in comparison with the disappearance of the penumbra in the continuum intensity maps. In particular, although the filaments are no longer visible in the continuum, the field is still horizontal and the Evershed flow has not completely vanished (see the area delimited by black arrows in continuum intensity, LP, and LOS velocity maps of Fig.~\ref{fig:fig_hinode1}). However, at the location of some of the disappearing filaments, blue-shifted patches are  %counter-Evershed flow seems to be 
detected (see the square in the panels b of Fig.~\ref{fig:fig_hinode1}). These could be attributed to blue shifts due to the 5-min oscillations, which cannot be removed in this Hinode dataset consisting of frames at three times only. Alternatively, they might be
related to the remnant of the upflows in the inner penumbra reported by
\citet{Rimmele1995}, provided that they seem to be linked to the horizontal magnetic field. Nevertheless, we cannot rule out that they could be interpreted as counter-Evershed flows (see \citealp{Murabito2016,SiuTapia2017,SiuTapia2018}).

\begin{figure*}
	\centering
	\includegraphics[scale=0.5,clip,trim=0 0 0 35]{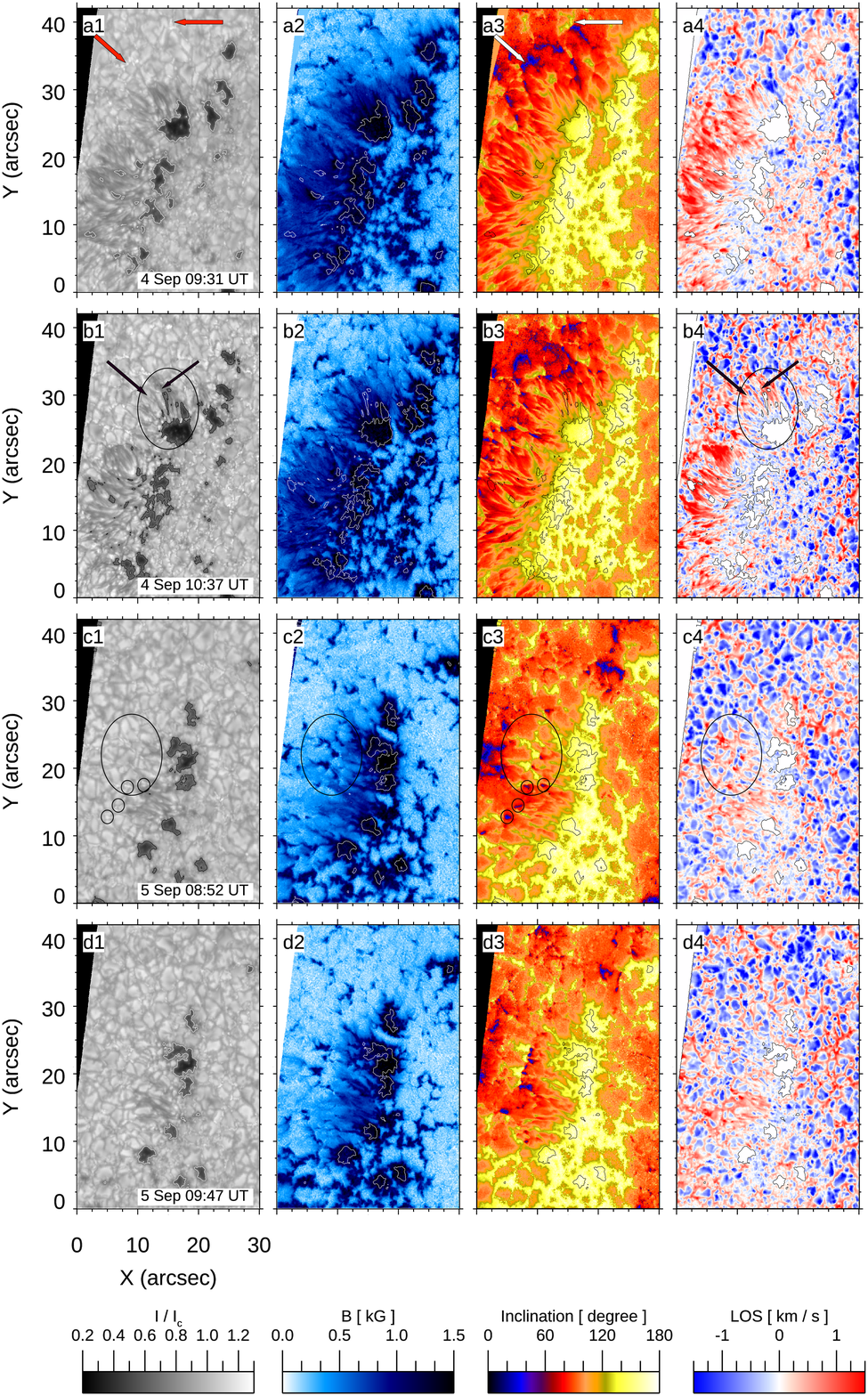}
	\caption{From left to right: Continuum intensity, magnetic field strength, magnetic field inclination, and LOS velocity maps from SST/CRISP observations at four representative times on September~4 and~5. Arrows and ovals mark features described in Sect.~3.2. }
	\label{fig:sst}
\end{figure*}

The CRISP instrument acquired data of the central part of the AR (i.e., similar to that shown in Fig.~\ref{fig:fig_hinode1}) for more than one hour on September~4. Figure~\ref{fig:sst} displays maps of continuum intensity (panels a1 and b1), magnetic field strength (panels a2 and b2), inclination angle (panels a3 and b3) and LOS velocity (panels a4 and b4) derived from CRISP data relevant to the observations acquired at 09:31 and 10:37~UT, respectively.

These data, which are characterized by a even higher spatial resolution than the ones from \textit{Hinode}/SP, allow us to study in more detail the penumbral grains and the patchy bright features, %\textcolor{red}{attenzione: in Hinode erano cose diverse, qui sono la stessa cosa?!?!}
in the inner part of the penumbra around the edge of the southern umbrae (see panels a1 and b1 of Fig.~\ref{fig:sst}). These penumbral grains, also visible in the \textit{Hinode}/SP data, exhibit both inward and outward motions. The magnetic field strength inside the small umbral cores is stronger than 1.2~kG and it does not show any appreciable changes after one hour, while the field decreases inside the penumbra (identified by the area between $I_{c}=0.7$ and $I_{c}=0.9$) reaching an average value of 0.8~kG (compare panels b1 and b2 of Fig.~\ref{fig:sst}).

The inclination maps show very well the fine structure of the penumbra, with inclination ranging between $80^{\circ}$ and $120^{\circ}$. A ring of patches with a field inclination of $20^{\circ} - 30^{\circ}$ encircles the northern side of the penumbra (see panel a3 of Fig.~\ref{fig:sst}, white arrows). These opposite polarity (respect with the small umbral cores) patches correspond to localized bright points in the continuum intensity map, as indicated by the red arrows in panel a1 of Fig.~\ref{fig:sst}).  
The limited FOV of the CRISP instrument does not allow us to say if other patches are present along the south-eastern side of the penumbra as well.
The LOS velocity map at 09:31~UT (panel a4 of Fig.~\ref{fig:sst}) shows the classical Evershed flow pattern inside the whole penumbra, with velocity values on average around $0.5 \,\mathrm{km \, s}^{-1}$ and maximum values around $2.7 \,\mathrm{km \, s}^{-1}$. %\textcolor{red}{perch\'e metti il meno nei valori della velocit\'a ???}
One hour later, a few filaments in the northern part of the penumbra start to disappear. Nevertheless, the Evershed flow pattern remains while penumbral filaments disappear (compare the regions indicated by the black arrows in panels b1 and b4 of Fig.~\ref{fig:sst}).

\begin{figure}
	\centering
	\includegraphics[scale=0.24,clip,trim=10 280 100 170]{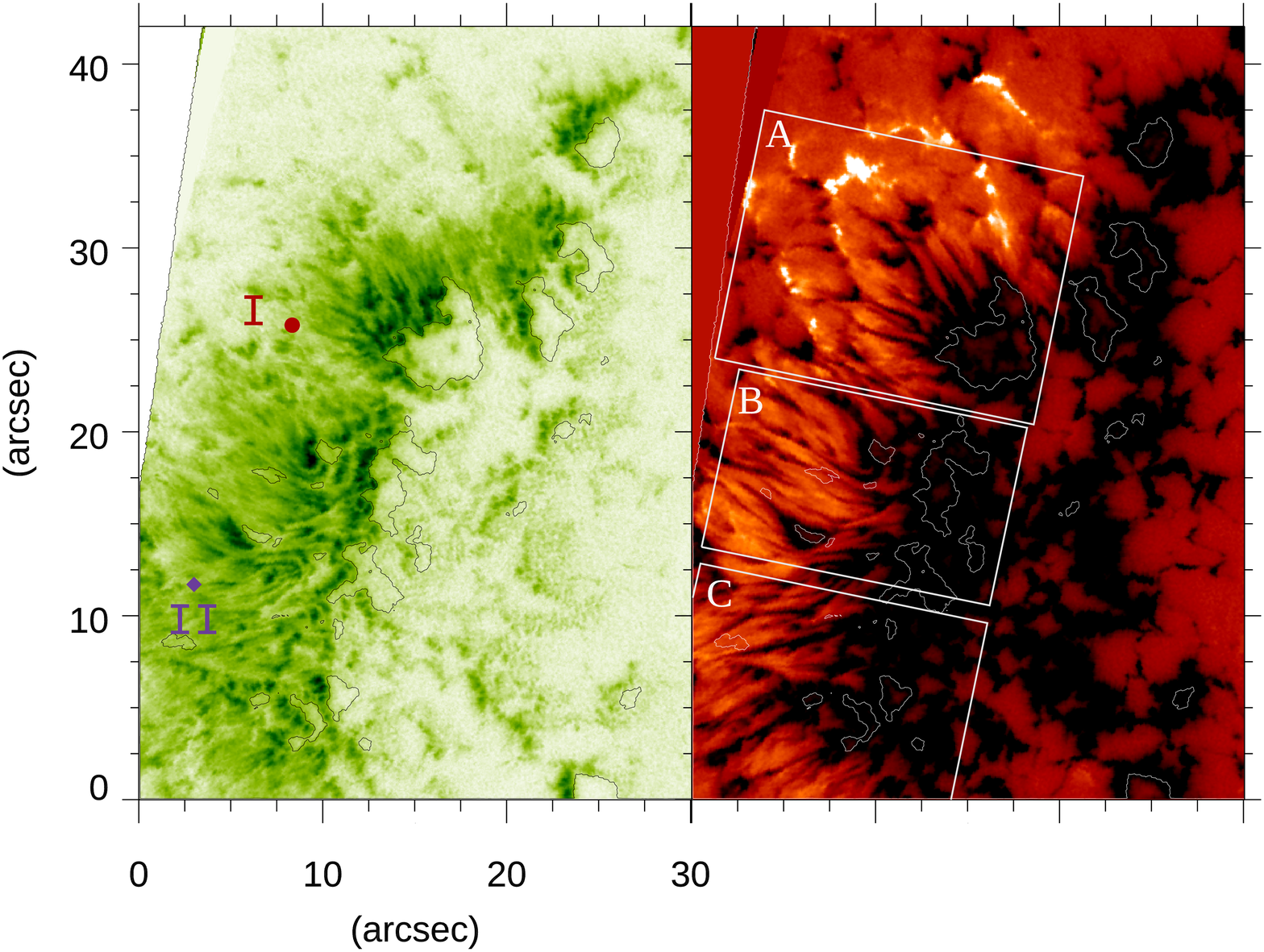}
	\includegraphics[scale=0.24,clip,trim=10 150 100 170]{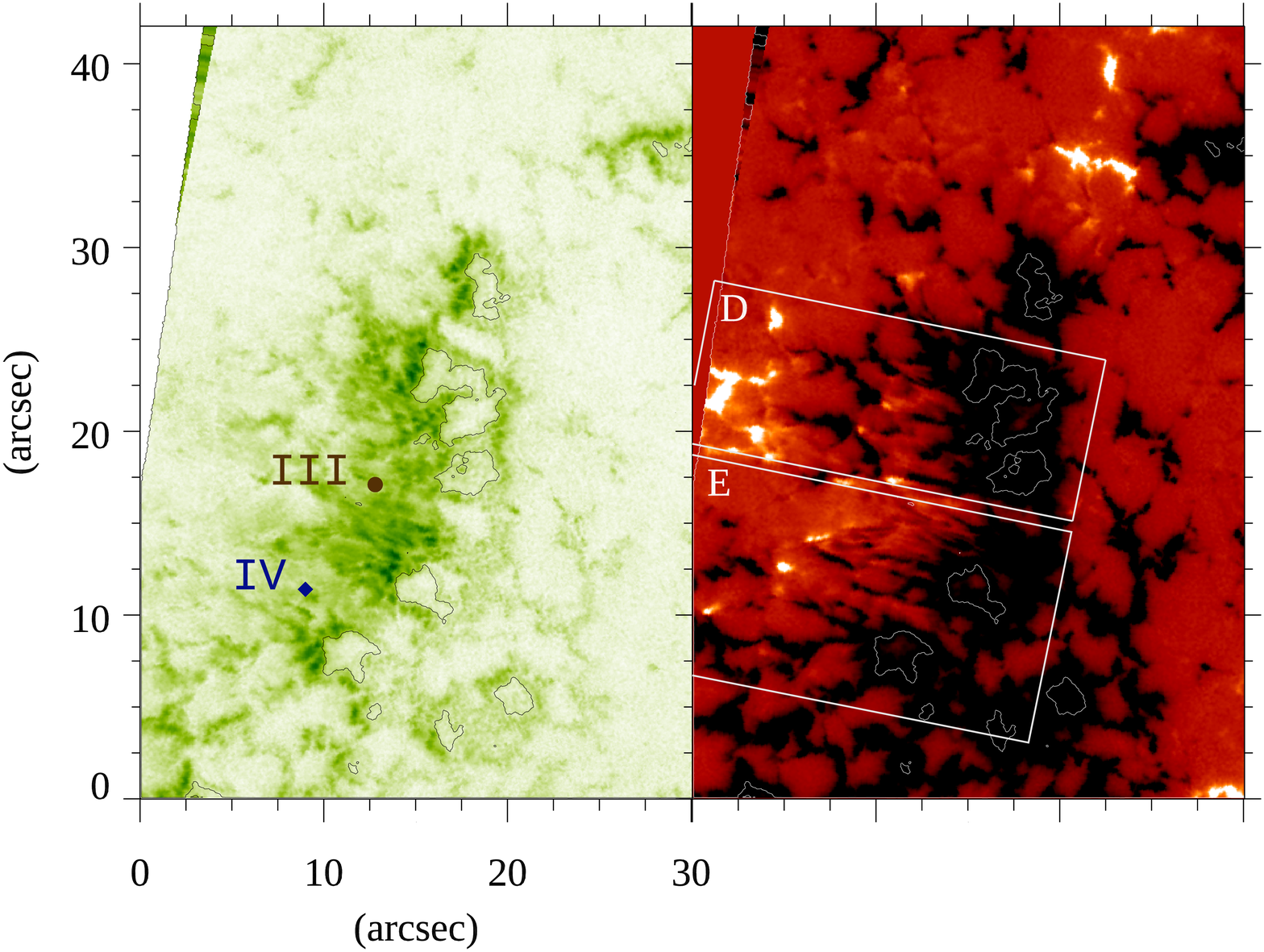}
	\caption{LP and CP maps from SST/CRISP data at 09:31~UT (top panels) and 08:52~UT (bottom panels) on September~4 and~5, respectively. The boxes drawn on the CP maps indicate where the magnetic flux evolution shown in Fig.~\ref{fig:sst_flux} has been computed. The LP and CP maps are scaled to $0.05 \,LP/I_{c}$ and $\pm 0.05 \,CP/I_{c}$, respectively. The coloured symbols (I-IV) in LP panels refer to locations considered in Fig.~\ref{fig:sst_profili1}. See Sect.~3.2 for more details.}
	\label{fig:new_figure}
\end{figure}

\begin{figure}[!h]
\centering
	\includegraphics[scale=0.24,clip,trim=1090 623 300 0]{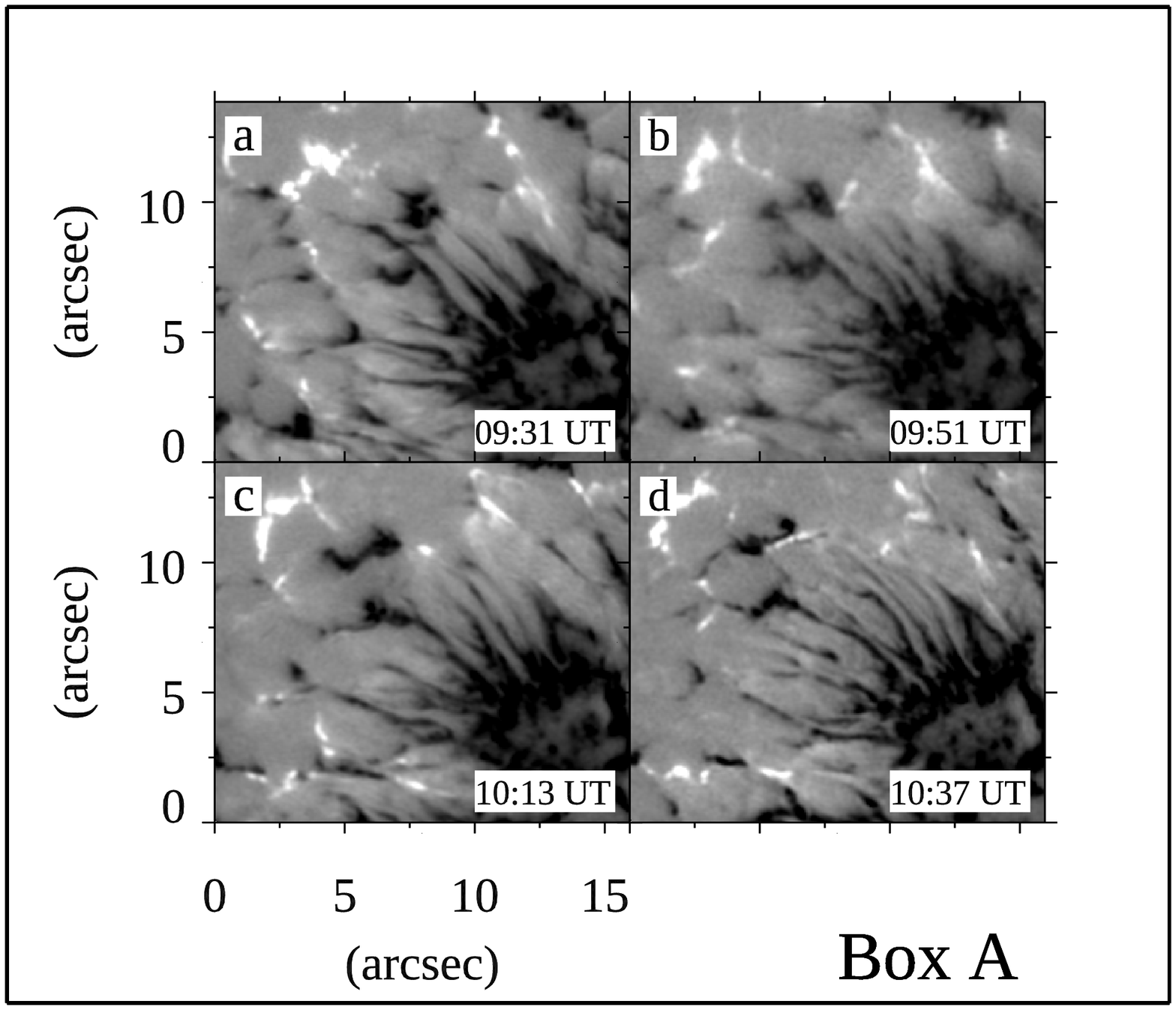}\\
	\includegraphics[scale=0.24,clip,trim=1090 80 300 35]{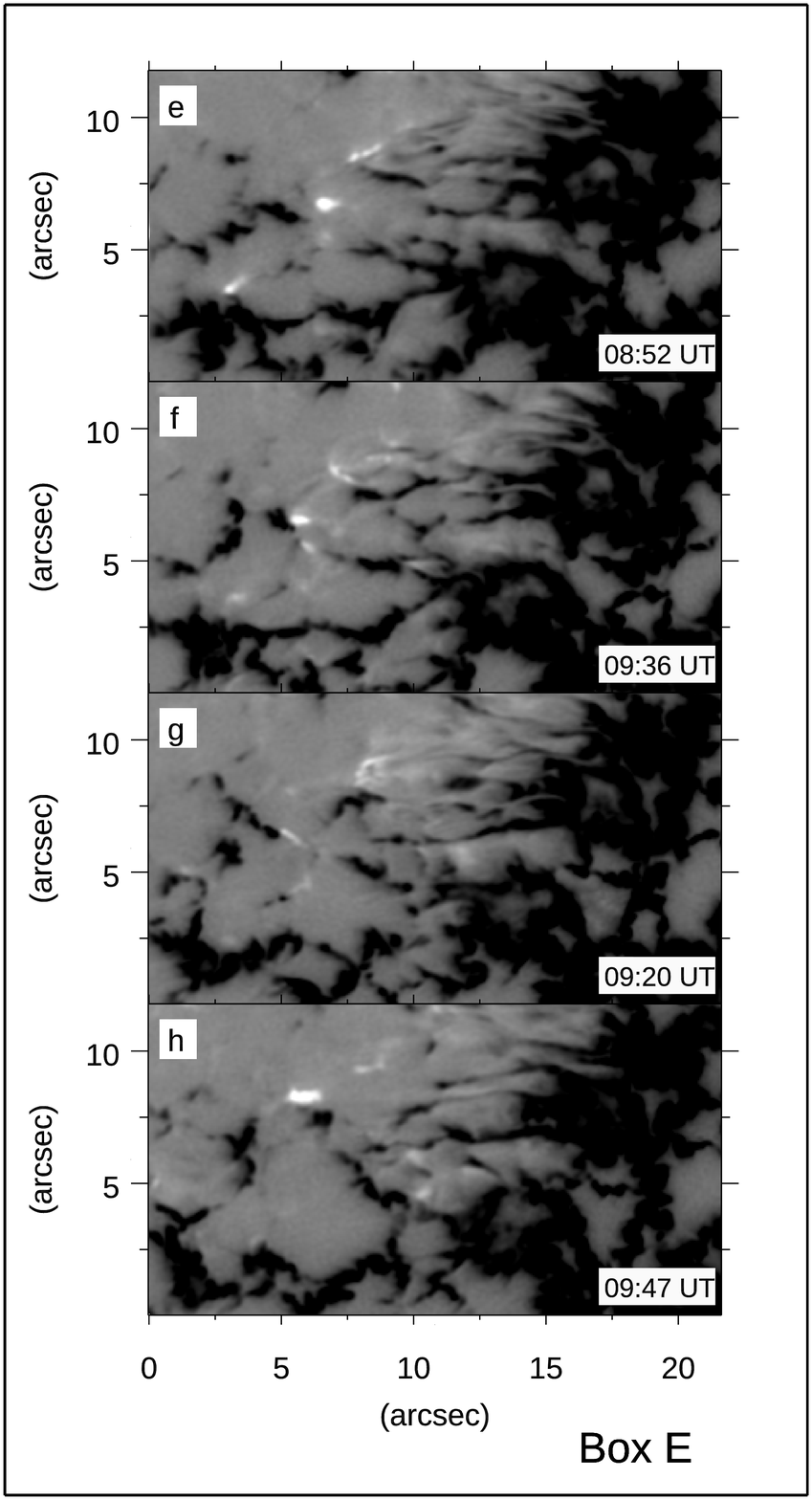}\\
	\includegraphics[scale=0.24,clip,trim=1090 220 300 34]{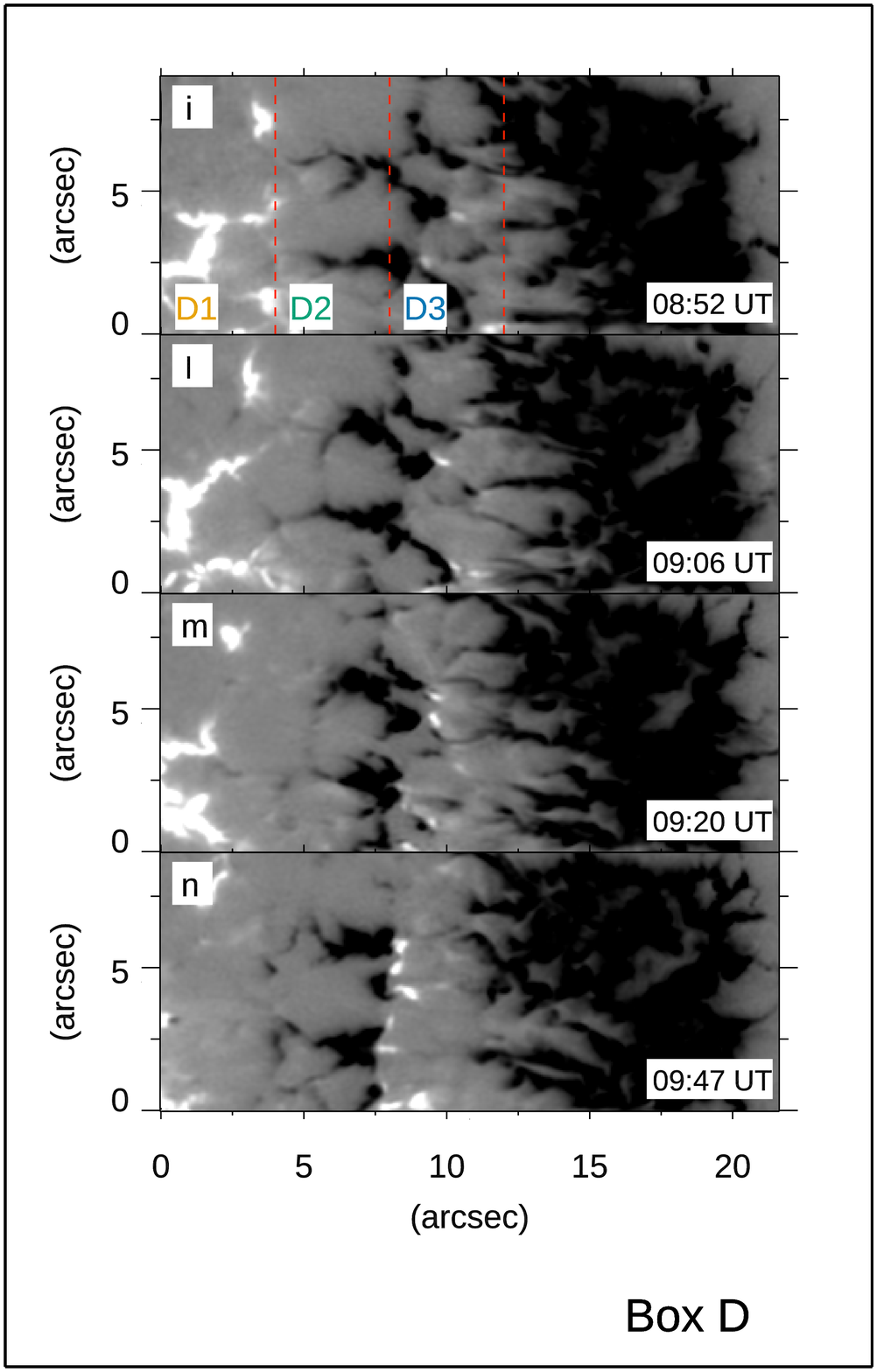}
	\caption{Zoomed CP maps of the box A (panels a-d), E (panels e-h) and D (panels i-n) from the SST/CRISP data at four representative times on September~4 and ~5. The boxes are indicated in Fig.~\ref{fig:new_figure}. All the CP maps are scaled to $\pm 0.05$ $CP/I_{c}$. The three red dashed lines in the first panel relevant to Box D indicate the three regions where the magnetic flux evolution reported in Fig.~\ref{fig:sst_flux_new} has been computed.
	\label{fig:sea}}
\end{figure}

\begin{figure*}
	\centering
	\includegraphics[scale=0.38,clip,trim=100 30 100 190]{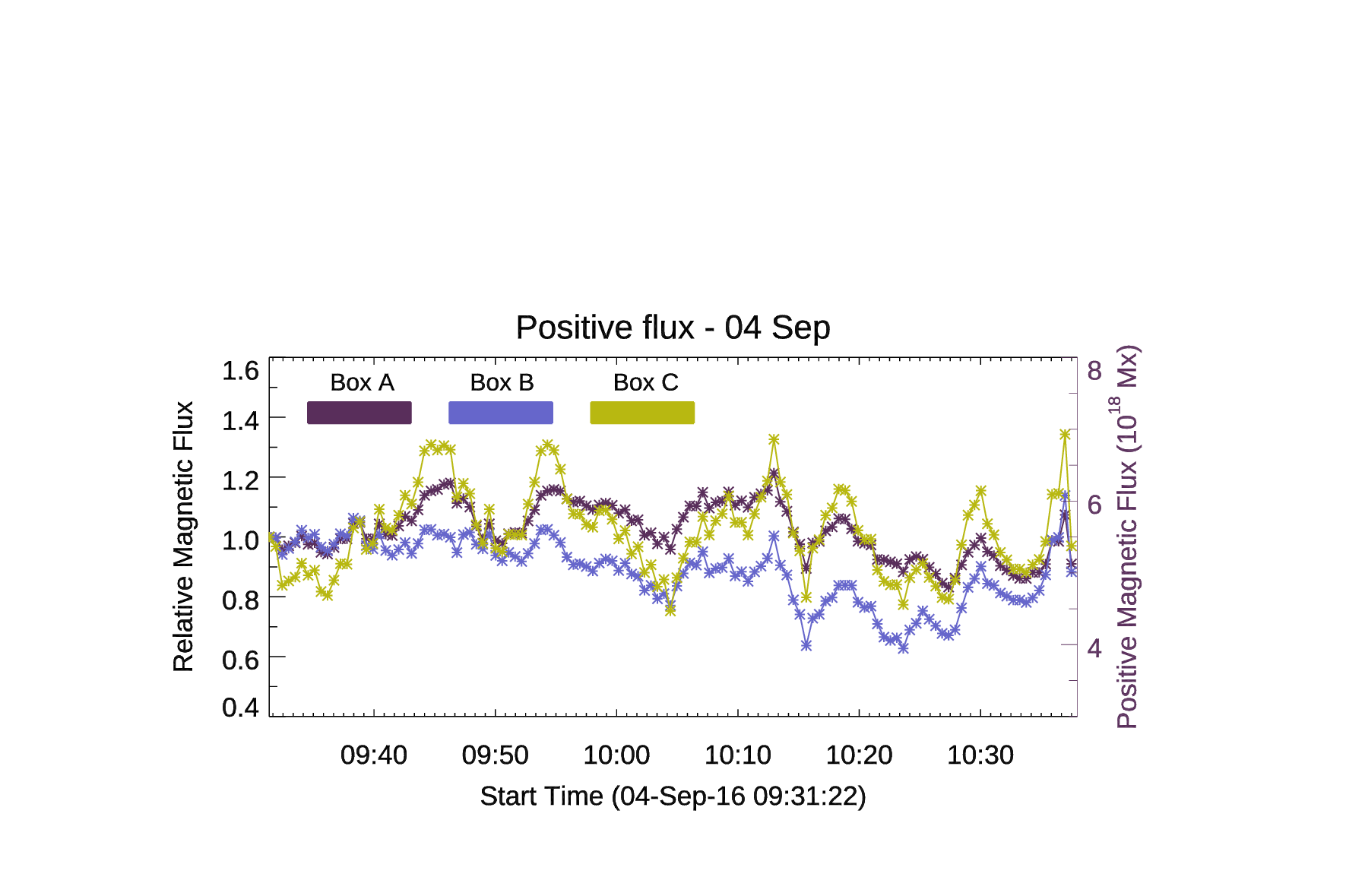}
	\includegraphics[scale=0.38,clip,trim=80 30 100 190]{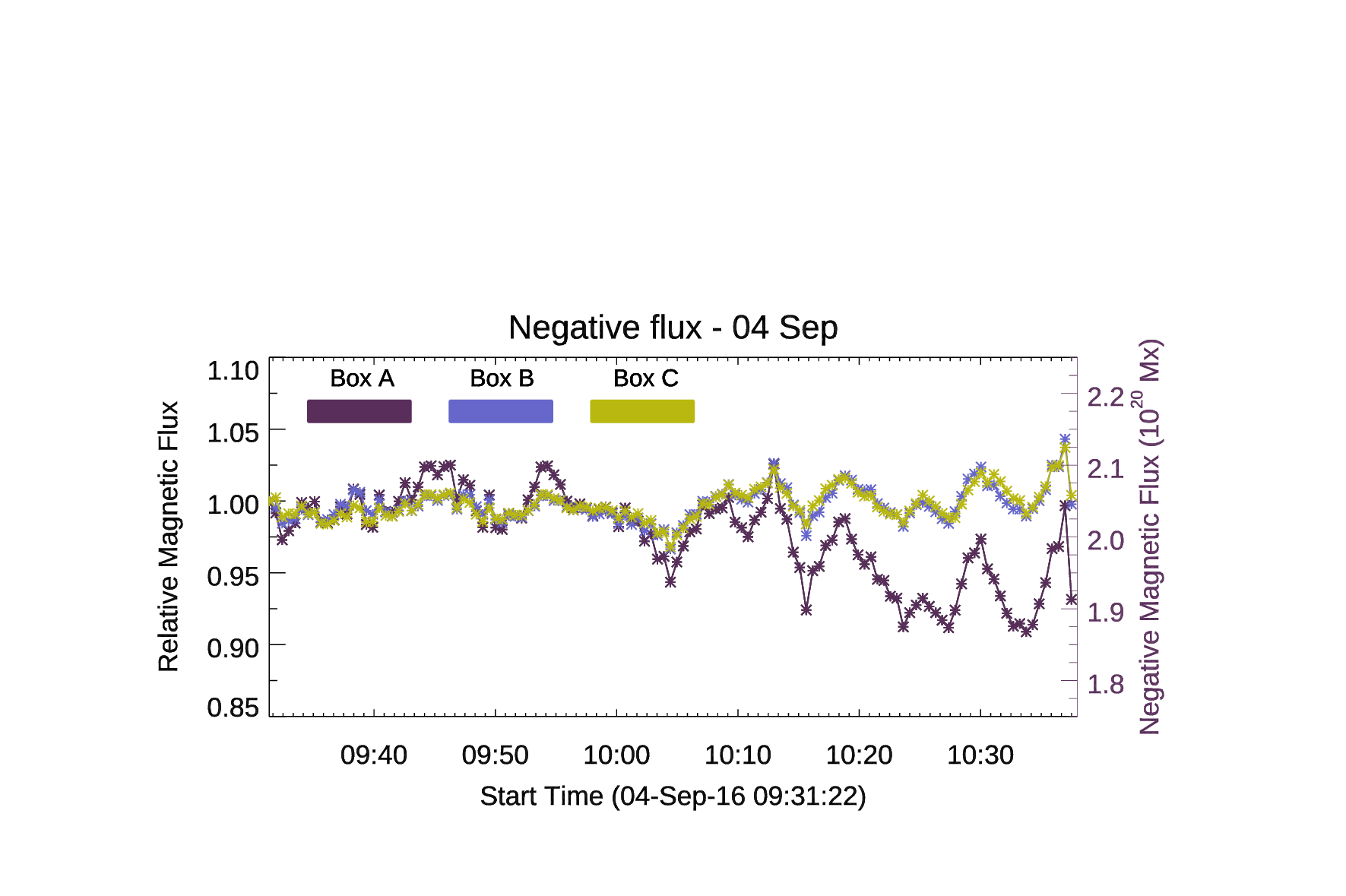}\\
	\includegraphics[scale=0.38,clip,trim=100 30 100 190]{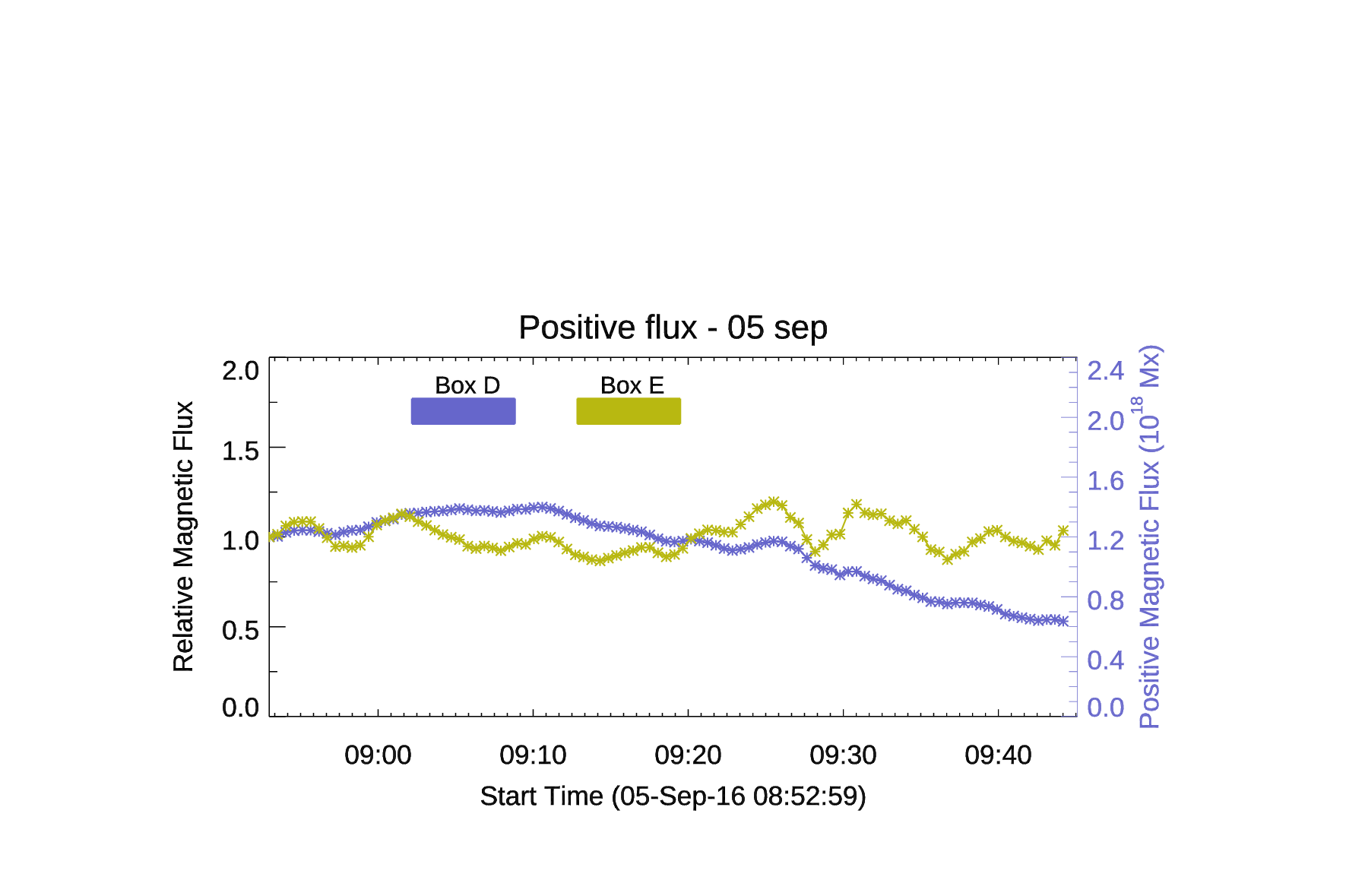}
	\includegraphics[scale=0.38,clip,trim=80 30 100 190]{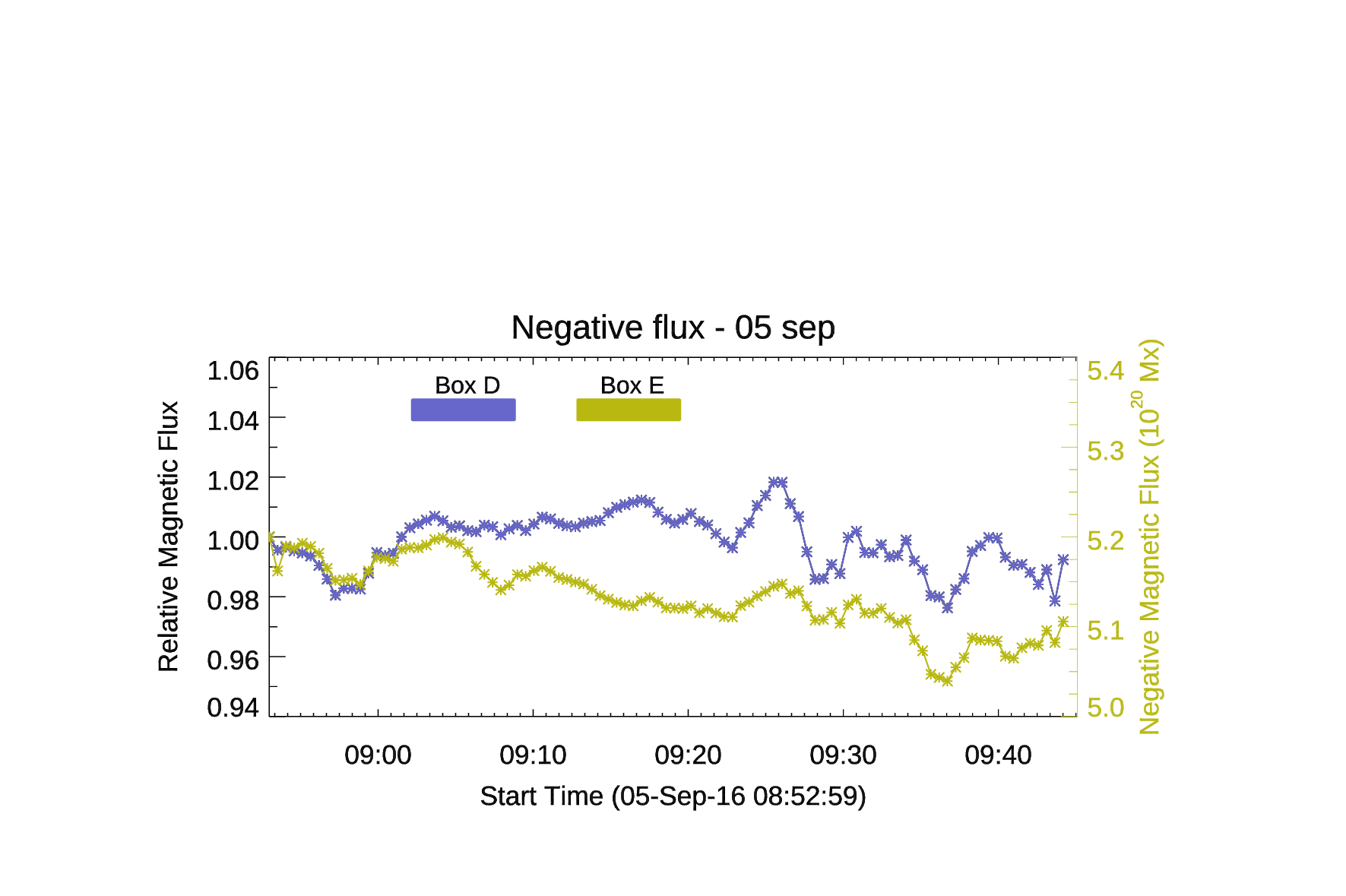}
	\caption{Evolution of the positive (left panels) and negative (right panels) magnetic flux inside the three boxes marked on the CP maps of Fig.~\ref{fig:new_figure} computed from the VFISV inversions of the SST/CRISP data on September~4 (top panels) and~5 (bottom panels). Symbols with different colours refer to the various analyzed boxes as specified in the legend.}  \label{fig:sst_flux}
\end{figure*}

\begin{figure*}
	\centering
	\includegraphics[scale=0.38,clip,trim=100 113 100 190]{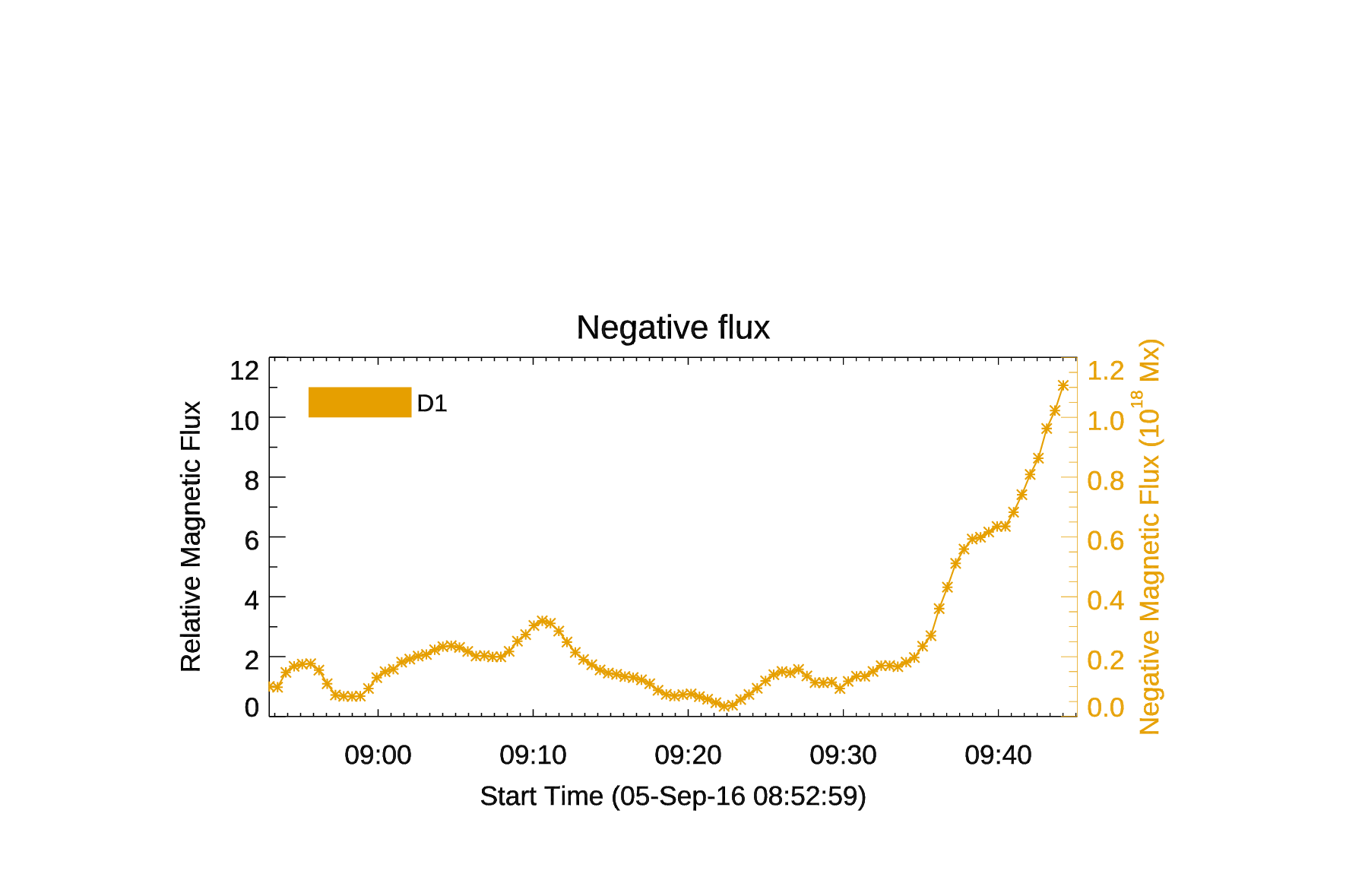}
	\includegraphics[scale=0.38,clip,trim=80 113 100 190]{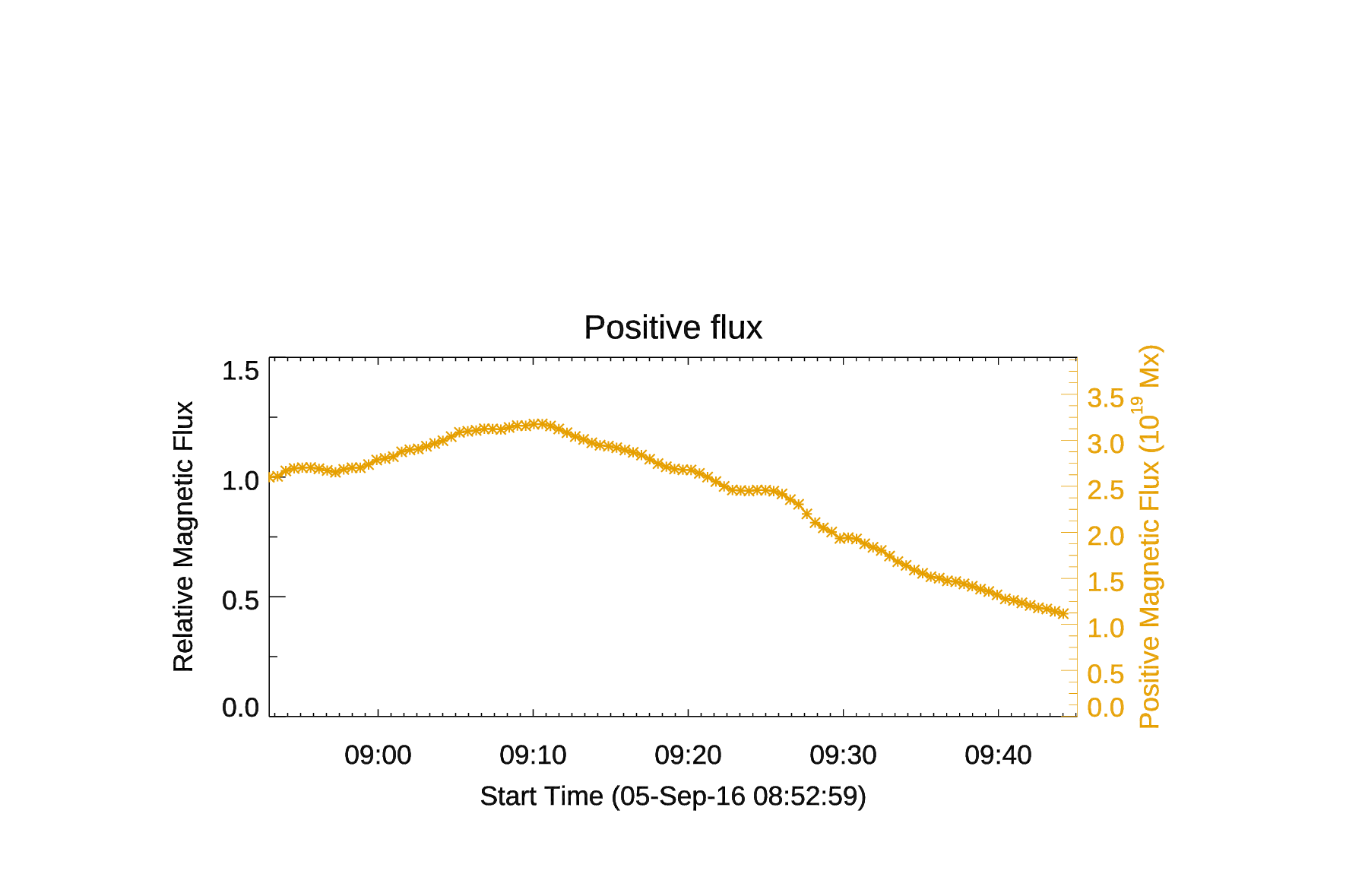}\\
	\includegraphics[scale=0.38,clip,trim=100 113 100 200]{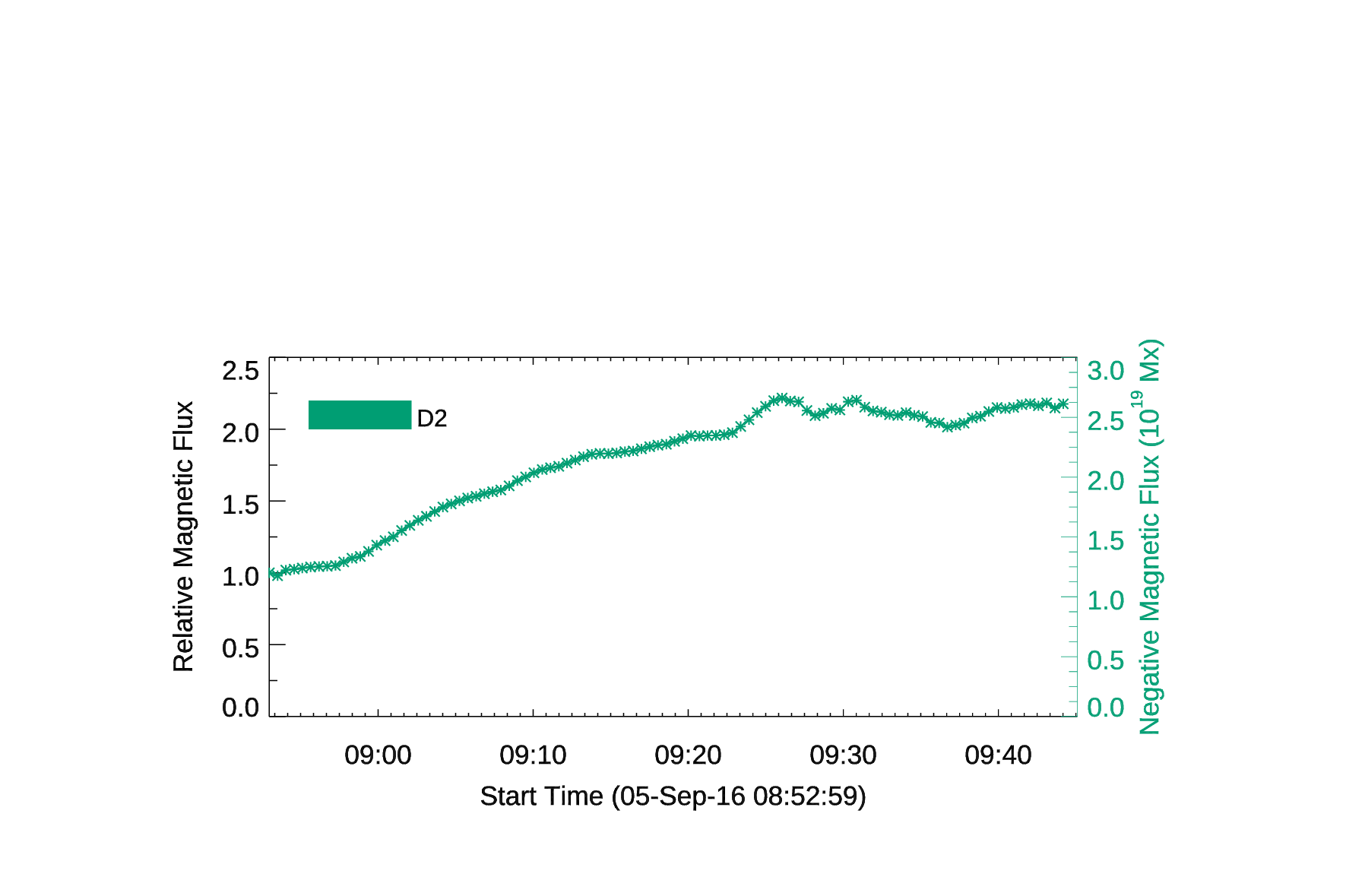}
	\includegraphics[scale=0.38,clip,trim=80 113 100 200]{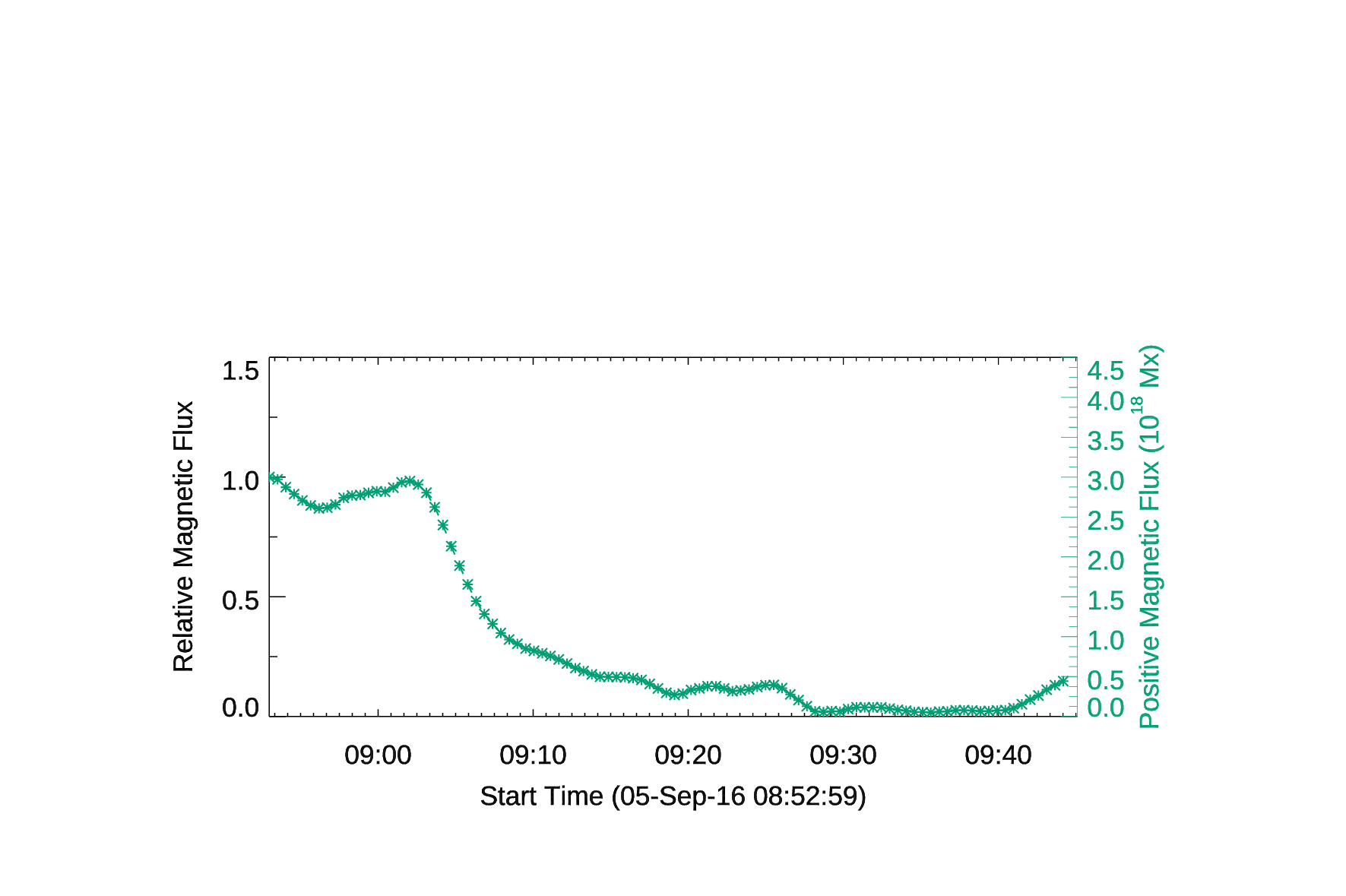}\\
	\includegraphics[scale=0.38,clip,trim=100 30 100 200]{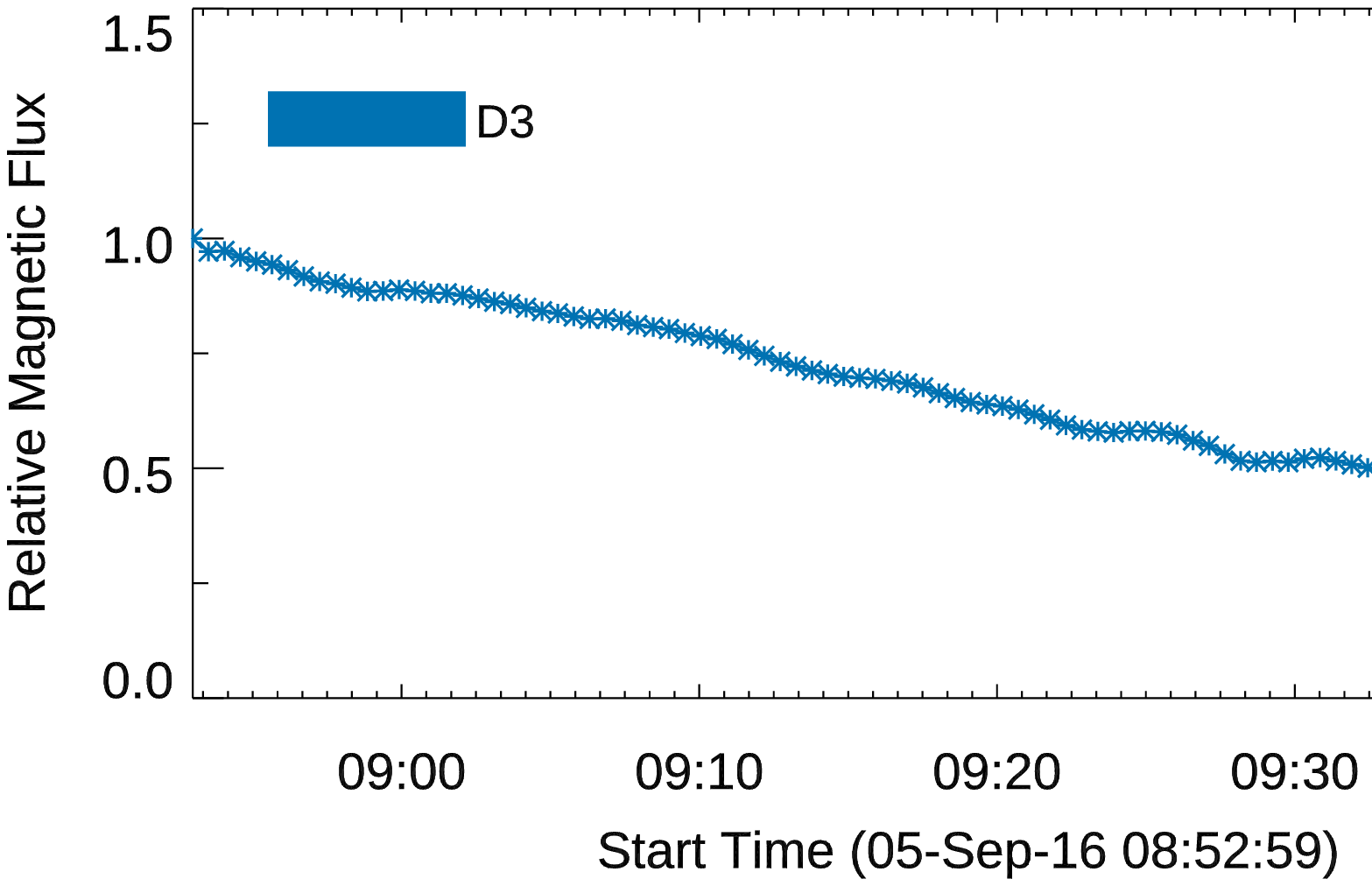}
	\includegraphics[scale=0.38,clip,trim=80 30 100 200]{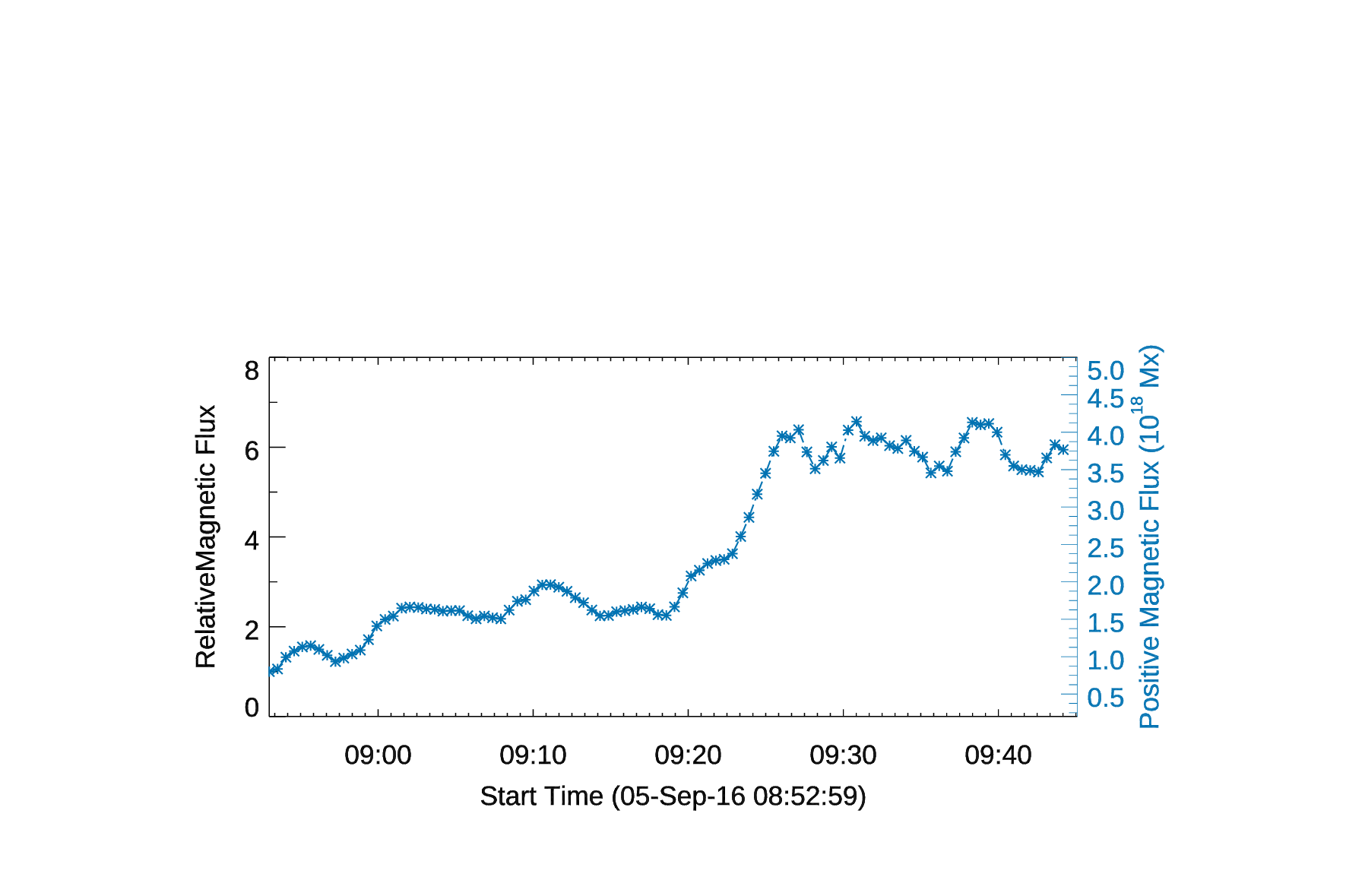}\\
	\caption{Evolution of the positive (left panels) and negative (right panels) magnetic flux inside the three regions D1, D2 and D3 marked on the CP map of Fig.~\ref{fig:new_figure} on September~5. Symbols with different colours refer to the various analyzed boxes as specified in the legend.}  \label{fig:sst_flux_new}
\end{figure*}

On September~5 (panels c and d of Fig.~\ref{fig:sst}) only few penumbral filaments are visible in the southern part of the system. In the first temporal frame, at 08:52~UT, the area located at the northern part of the remaining penumbral sector displays elongated and stretched granules (see the area indicated by the ovals in panels c of Fig.~\ref{fig:sst}).  
Panels c2 and c3 of Fig.~\ref{fig:sst} reveal a magnetic field with a sea-serpent-like configuration \citep{Dalda&Bellot2008}. We clearly see large patches of opposite polarity, for instance at X,Y=[5\arcsec,20\arcsec] and [25\arcsec,35\arcsec]. Moreover, at the edge of the penumbral filaments in the inclination map there are patches of opposite polarity with respect to the umbral field (see the small ovals in panel c1 and c3 of Fig.~\ref{fig:sst}). In less than one hour, at 09:47~UT (panels d of Fig.~\ref{fig:sst}), the remaining penumbral filaments become thinner and the large-scale opposite polarity patches at the edge of them shrink. Conversely, the structure inside the oval is characterized by coexisting opposite polarity features (see panels c2 and c3 of Fig.~\ref{fig:sst}). This will be further investigated in the following (see Fig.~\ref{fig:sea}).\\ In order to investigate the dynamics of the fine penumbral structures involved in the penumbral decay, we benefitted from polarization maps derived from CRISP measurements, shown in Fig.~\ref{fig:new_figure}. In the analyzed region on September~4 horizontal magnetic fields and a large number of MMFs are detected in the time series of CP and LP maps. In particular, the movie available in the online material, relevant to LP map from September 4 data (top-left panel of Fig.~\ref{fig:new_figure}), displays that the area of the LP signal cospatial to the region where the penumbral filaments are disappearing (near the red filled circle) shrinks along time, although it does not completely vanishes.\\
The zoomed panels (a-d) of Fig.~\ref{fig:sea}, corresponding to the box A in the CP map from September 4 data (top-right panel of Fig.~\ref{fig:new_figure}), show the interaction between the penumbral field and opposite polarity patches that could lead to cancel out part of the penumbra. In order to study this aspect, we examined the longitudinal magnetic flux evolution as retrieved from the VFISV inversions of the whole SST/CRISP data. The plots reported in the top panels of Fig.~\ref{fig:sst_flux} present the relative positive (left panel) and negative (right panel) magnetic flux inside the three boxes drawn over the CP map of the top panels of Fig.~\ref{fig:new_figure}. It is evident that the negative magnetic flux decreases by about $0.2\times 10^{20}$~Mx inside the box A only (dark violet curve, right panel), where the penumbral filaments disappear. Conversely, the bluish (box B) and yellowish (box C) curves remained almost constant with time. On the whole, the plot reported in the left panel of Fig.~\ref{fig:sst_flux} reveals that a significant positive magnetic flux inside the box~A remains almost constant, about $5.5 \times 10^{18}$~Mx, twice as large as the average flux content in box~B ($2.0 \times 10^{18}$~Mx) and~C ($2.3 \times 10^{18}$~Mx). This result is in line with the the idea that part of the penumbral field, characterized by a negative polarity, is canceled out by the interaction with the positive polarity patches seen in the CP movie also available in the online material. Concerning the southern part of the region, the CP movie shows that there is no significant concentration of opposite polarity patches interacting with the penumbral field. Indeed, as already anticipated, in box~B and box~C there is no decrease of the negative magnetic flux producing the disappearance of the penumbra, even if the positive flux remains almost constant.\\
We used a similar approach to highlight how the penumbral decay process proceeds on September~5 (see bottom panels of Fig.~\ref{fig:new_figure}). First, from the CP map, we note that the elongated granules, located at the northern part of the remaining penumbral sector and characterized by blueshifted elongated patches in the LOS velocity maps (see the panels c1 and c4 of Fig.~\ref{fig:sst}), are exactly at the right edge of the sea-serpent-like configuration of the field (see the CP map in the bottom-right panel of Fig.~\ref{fig:new_figure}), which is also visible in the inclination maps (panels c3 and d3 of Fig.~\ref{fig:sst}). Although the penumbra has almost disappeared, LP signal is still detected around the small umbral cores where the filaments disappeared and elongated granules appear (see LP map in the bottom-left panel of Fig.~\ref{fig:new_figure}). The box~E is the only area where some penumbral filaments are still present, while decreasing in size with time. The zoomed panels (e-h) of Fig.~\ref{fig:sea} illustrate the evolution of this area within the one hour of SST/CRISP observations. As reported for September~4, numerous MMFs can be detected. The bottom panels of Fig.~\ref{fig:sst_flux} contain the plots for the positive/negative magnetic flux evolution inside the two boxes D and E marked over the CP map in bottom-right panel of Fig.~\ref{fig:new_figure}. According to this figure, the negative magnetic flux (right panel) in both boxes shows a decrease even if with different trends. Instead, the positive fluxes (left panel) have a different behavior. In fact, while in the box~E the positive flux remains constant with time (yellowish curves), this is not the case for the box~D (bluish curves). These results are consistent with the idea of the interaction between type III MMFs and the penumbral field.\\
To disentangle the various magnetic contributions occurring inside the Box D, we also studied the evolution of the positive (negative) flux variation inside three regions marked in panel (i) of Fig.~\ref{fig:sea}. In particular, we considered the D1 region between X=0\arcsec and X=4\arcsec, the D2 region from X=4\arcsec to X=8\arcsec and the D3 region between X=8\arcsec and X=12\arcsec. \textbf{Figure~\ref{fig:sst_flux_new}} displays the variation of the magnetic flux inside these three regions. The plots concerning the region D1 (first row of \textbf{Fig.~\ref{fig:sst_flux_new}}) show a decrease of positive magnetic flux and an increase of the negative one. It is worth noting that the positive flux could decrease because the positive polarity patches seen in this region were not longer included in the FOV. However, these changes in the evolution of the magnetic flux do not have relation with the penumbra and its disappearance. The other two regions (D2 and D3) include the area where the small-scale sea-serpent magnetic configuration is forming at about $2\arcsec - 3\arcsec$ away along the prolongation of the penumbral filaments. A constant increase of the negative magnetic flux is detected for the area inside D2, while this region exhibits a decrease of the positive magnetic flux, although with different trends (see second row in \textbf{Fig.~\ref{fig:sst_flux_new}}). An opposite behaviour is found for the region D3. Indeed, a constant decrease of the negative magnetic flux is associated with a positive magnetic flux increase, even with different rates (see third row in \textbf{Fig.~\ref{fig:sst_flux_new}}). In particular, the emergence of the positive magnetic flux seen in this latter plot, due to an emerging region moving toward the negative polarity, cancels part of the pre-existing negative polarity flux. Unfortunately, we do not have data acquiring the evolution of this particular structure. However, 
%\textbf{The} noteworthy behaviour, already noticed in the inclination maps reported in Fig.~\ref{fig:sst} \textbf{and in the plots concerning the D2 and D3 areas}, concerns the formation of the small-scale sea-serpent configuration at about $2\arcsec - 3\arcsec$ away along the prolongation of the penumbral filaments. 
in the zoomed panels (i-n) of Fig.~\ref{fig:sea}, we observe the separation of type III MMFs that converge towards opposite polarity patches, forming in about one hour the complex, well-outlined structure like a magnetic wall, in line with the evolution of the flux reported above. 

\begin{figure}[!h]
\centering
   \includegraphics[scale=0.5,clip,trim=40 70 410 30]{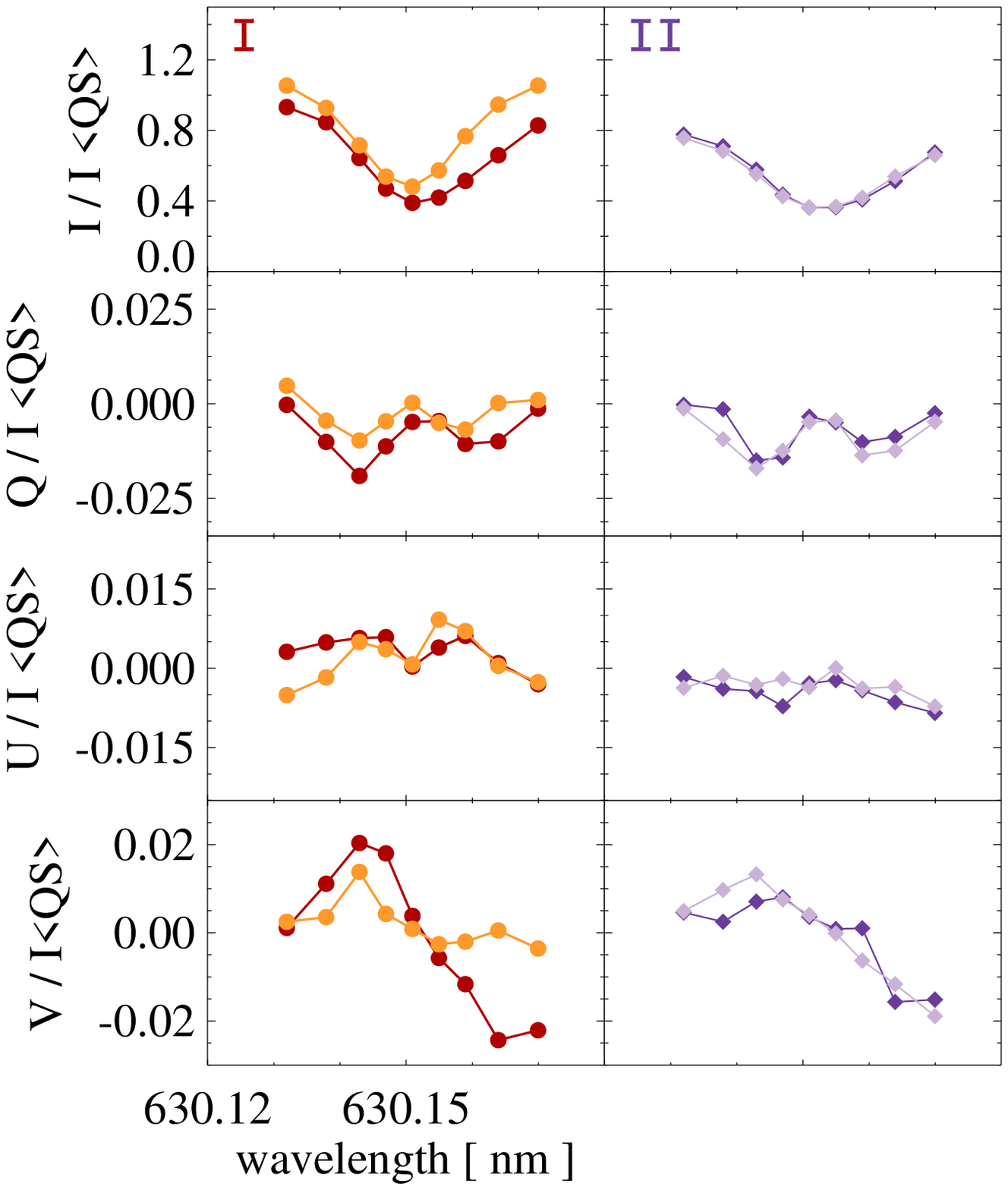}
     \includegraphics[scale=0.5,clip,trim=40 70 410 30]{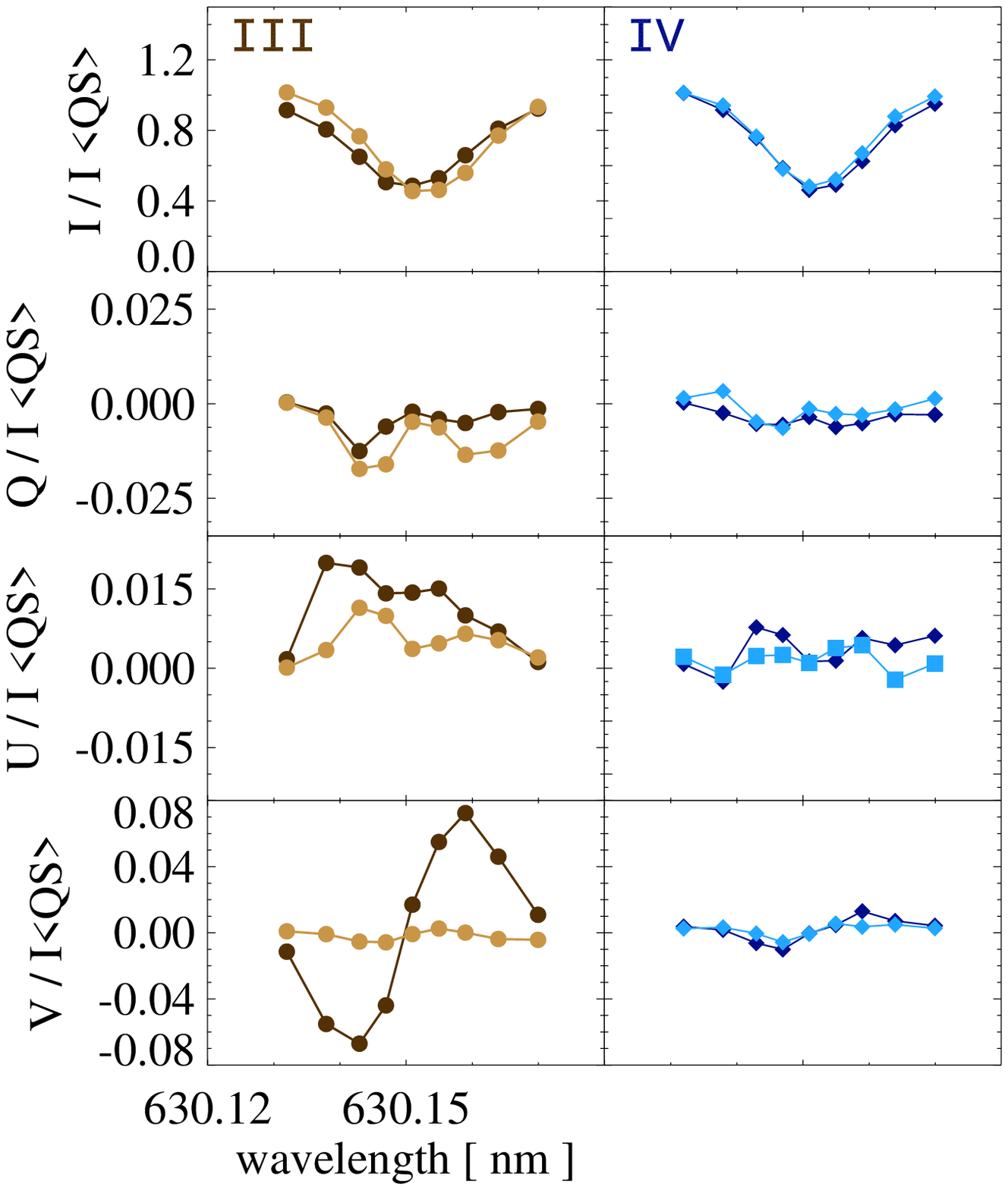}  
    \caption{Stokes profiles observed in the pixels (I-IV) marked with coloured filled symbols in the LP maps of Fig.~\ref{fig:new_figure} from SST/CRISP observations on September~4 and~5. From top to bottom, we show the four I, Q, U and V Stokes profiles. The colours indicate initial (dark colours symbols) and final (light colours symbols) times of the observations.}
    \label{fig:sst_profili1}
\end{figure}

In Fig.~\ref{fig:sst_profili1} we present examples of the observed Stokes profiles in the pixels  corresponding to the coloured symbols (I-IV) drawn on the LP maps of Fig.~\ref{fig:new_figure}. The initial time is represented with a darker colour, while the ending time with a lighter colour. The red and blue symbols (I and II) on the LP map on September~4, are located inside the boxes~A and~C, respectively. In the case of the red symbol (i.e., position marked with I), we note that the Stokes $V$ signal slightly decreases with time while the penumbral filaments disappear in that location (see panel b1 of Fig.~\ref{fig:sst}). At the same time, the Stokes $Q$ and $U$ profiles do not show significant variations. Conversely, the location labelled with the violet diamond symbol (i.e., position marked with II) displays the almost unchanged Stokes $Q$, $U$ and $V$ profiles. Similarly, the brown and blue coloured symbols (i.e., positions labelled with III and IV, respectively) identify special positions, both inside the box C, on 5 September (see the LP map on the bottom-left panel of Fig.~\ref{fig:new_figure}). The main difference is that one of these pixel positions (the brown one) is located in the area where the penumbra is disappearing and where a large number of MMFs are seen. In this particular case the most important signature results in the decrease of the Stokes $V$ profiles, while the Stokes $Q$ and $U$ profiles do not display considerable changes. On the other hand, the blue symbol, which resides in a ``quiet'' area, display all the Stokes profiles unchanged, meaning that probably no opposite (positive) polarity patches interact.

\begin{figure} 
\centering
\includegraphics[scale=0.2,clip,trim=0 0 0 0]{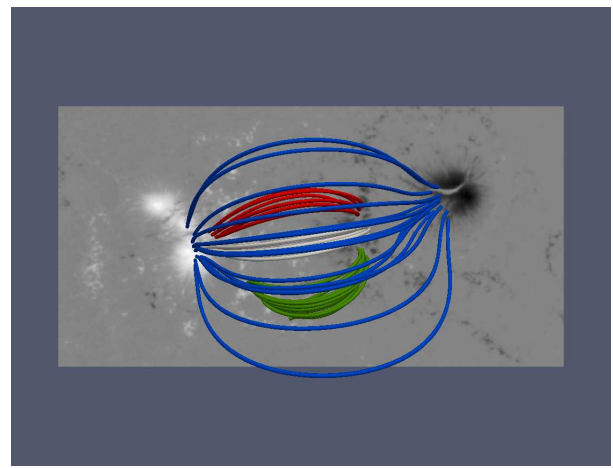}
    \caption{Potential field extrapolation obtained using SDO/HMI LOS magntogram taken on September~3 at 08:48~UT as boundary condition.}
    \label{fig:estrapolazione}
\end{figure}

\section{Discussion and Conclusion}

Multi-instrument photospheric observations have allowed us to investigate the decay process of a penumbra in the central part of the AR NOAA~12585.

Thanks to the synoptic SDO/HMI observations, we have been able to follow the entire evolution of this AR from its appearance on the East solar limb on 2016 September~1 till the disappearance of the penumbral areas in its central part. During its evolution (from September~1 to~3), the AR first exhibits a $\beta$ magnetic configuration that changes with time becoming more complex and showing the formation of magnetic structures between the two main polarities. 
During this time interval, two not fully developed sunspots form in the central region. In this context, we remark a noteworthy aspect about the formation of the penumbra of these two structures: i.e., for both the structures, the penumbra forms only in the region towards the opposite polarity, whereas the area toward the leading spot does not show any penumbral sector. This finding, i.e. the penumbrae do not occur between nearby spots of the same polarity, was already noted by \citet{Kunzel1969}.

The formation of a penumbra towards the opposite polarity in a flux emergence region occurs in apparent contrast with the observations of \citet{Schlichenmaier10,Schlichenmaier12} and \citet{Rezai12}. They found that during the penumbra formation process, penumbral filaments initially do not form in the side facing to the opposite polarity, where the ongoing emerging flux does not allow a stable penumbra to form. However, \citet{Murabito17} found that the penumbra forms also on the side toward the opposite polarity, i.e. the flux emergence region. Even considering the photospheric evolution of bipolar ARs with the simplest magnetic configuration, i.e. $\beta$ type, a preferred penumbral formation side was not found so far \citep{Murabito18}. On the other hand, this interpretation assumes that the penumbra formation is mainly linked to the emergence of the magnetic flux from the convection zone. Taking into account the scenario proposed by \citet{Romano13, Romano14, Romano20}, the penumbra settlement takes place after the magnetic flux emergence, when the magnetic canopy, already present in the upper layers of the solar atmosphere, changes the inclination of its magnetic field, reaching the photosphere and forming the penumbra. In this case, the preferred penumbral formation side does not depend on the location of the emerging magnetic bipole, but on the overlying magnetic field configuration and the magnetic connectivity. In this light, we performed a potential field extrapolation using the SDO/HMI LOS magnetogram taken on September~3 at 08:48~UT as boundary condition (see Fig.~\ref{fig:estrapolazione}). The potential field shows the connectivities of the central part of the AR with some positive and diffused polarities located eastwards (see the red, white and green loop systems in Fig.~\ref{fig:estrapolazione}). No positive polarities are visible near the leading polarity, therefore no field lines fill the region between the negative polarities of the central part of the AR and the main negative spot. The magnetic field configuration derived from our extrapolation of the HMI data suggests that the penumbra may not form without the presence of the overlying magnetic canopy.

With regard to the penumbal decay, we presented high-resolution measurements available on September~4 and~5 outlining the properties of vector magnetic field and LOS velocity during the disappearance of the different penumbral sectors in the central region of the AR. At that time, the penumbra looked like an orphan penumbra \citep{Zirin&Wang1991,Jurcak2014, Zuccarello2014}.

Although these high-resolution observations only cover a few short time intervals of the processes at work, we have found a number of remarkable and coherent aspects that could help in the understanding of the complex and poorly studied process of penumbral decay. 

The main results of our study can be summarized as follow:
\begin{enumerate}
    \item The negative magnetic flux of the central region decreases from September~3 to~5 with an average rate of $2 \times 10^{16} \,\mathrm{Mx \, s}^{-1}$ (corresponding to $1.7 \times 10^{21} \,\mathrm{Mx \, day}^{-1}$), till the penumbra completely disappears and only some pores survive. 
    \item Before the penumbra disappears, opposite polarity patches in the inclination maps are visible. These patches are associated with penumbral bright points in the continuum intensity.  
    \item Once the penumbral filaments disappear in the continuum intensity, the Evershed flow does not immediately cease to be, but it is only reduced. Furthermore, the area involved by the Evershed flow displays strong LP signals indicating that the field is not completely vertical. As a consequence, after the penumbra disappears, it is not immediately replaced by the granulation and the convective velocity field pattern. %Counter-Evershed flow occurs in some locations.
    \item The presence of MMFs, progressively located near the sectors which starts to disappear, as well as the magnetic flux evolution suggest that the penumbral magnetic flux may be removed by these features. Furthermore, the analysis of the Stokes profiles where the penumbra is disappearing displays a decrease of the Stokes $V$ profiles and unchanged Stokes $U$ and $Q$ profiles. This is in line with the idea of the flux removal due to the interaction between the negative penumbral field and opposite polarity features, as detected.
    \item A small sea-serpent-like magnetic configuration of the magnetic field results from the detached type III MMFs converging toward opposite polarity patches as the penumbral filaments disappeared.
\end{enumerate}

The reported decay rate ($1.7 \times 10^{21} \,\mathrm{Mx \, day}^{-1}$) is one order of magnitude greater than that found by \citet{Den07}, confirming that decay we have observed is faster in comparison with previous observed cases. However, we cannot neglect that \citet{Den07} have reported the decay rate of a regular sunspot with umbra and a complete penumbra, while we consider the decay process of a not fully developed sunspot embedded in a more complex magnetic configuration. On the other hand, the decay process simulated in \citet{Rempel2015} for a naked spot reports a steady decay of about $10^{21} \,\mathrm{Mx \, day}^{-1}$, in good agreement with our observed decay rate. Our decay rate measurements is also in line with that reported in \citet{Benko18}, taking into account the appropriate corrections.

Concerning the large-scale opposite polarity patches, we can surely link them to the penumbra formation. Indeed, \citet{Romano13,Romano14} found that, during the formation process of a sunspot penumbra, opposite polarity patches with respect to the parent umbra move radially outward. They interpreted these results as the displacements of the footpoints of the field lines stretching and returning to the photosphere in order to form the penumbra from the inclined magnetic field. In our case, the opposite polarity patches are found also before the disappearance of the penumbra, around the penumbral filaments and associated with localized bright points. Furthermore, in \citet{Murabito2016} the opposite polarity features found around the umbra before the penumbra forms, migrate outward forming a ring that moves away from the spot with time. They suggested that this ring moves presumably driven by the moat flow.
Bright features moving in the outer penumbra were also detected in the \textit{G}-band during the decay of sunspots by \citet{Kubo2008a}. Their interpretation supports the idea that subsurface upwelling and diverging flows can destabilize the spot, linking MMFs and convective motions. 
In this perspective, further work should be done to better investigate this aspect and disentangle whether an analogous process happens during the disappearance stage of the penumbra or this signature suggests a role of convection motions in eroding the footpoints of sunspot.

The relationship between the presence of Evershed flow and sunspot decay was studied by \citet{Bellot08}, who suggested that the penumbra at photospheric level disappears when the penumbral field lines that no longer carry strong Evershed flows rise to the chromosphere. Our results display that the Evershed flow and horizontal fields do not completely vanish after the disappearance of penumbral filaments in the continuum. Therefore, we argue that the Evershed flow persists as long as the horizontal magnetic field lies in the photosphere, although the strength of the magnetic field is not able to keep the reduced heating transport typical of the sunspot penumbra. In addition, blue shifted patches have been detected at the location where penumbral filaments are disappearing. These might be the first indication of reappearance of granulation, or they can be the related to the remnant of the upflows in the inner penumbra \citep{Rimmele1995} linked to the horizontal fields. We do not rule out that these blue-shifted patches might be related to counter-Evershed flows,%. counter-Evershed flow can be observed in some locations, 
as in turn observed by \citet{Murabito2016,Murabito18} during penumbra formation and by \citet{SiuTapia2017,SiuTapia2018} in a well-developed sunspot penumbrae. At certain locations, it may even reverse, like during penumbra formation, suggesting an inverse process in temporal sense between penumbral formation and decay. Granulation as well as the convective velocity field pattern take a longer time to reappear as the penumbra disappears in the continuum intensity.
In broader perspective, from the SST/CRISP observations, it was possible to highlight that the penumbra does not decay as whole, but different penumbral sectors disappear progressively. Indeed, we have caught the disappearance of two sectors of penumbra, one on the north-eastern side and the other one in the south-eastern side of the region. In such a perspective, the magnetic flux evolution, derived from the spectropolarimetric VFISV inversions of the whole SST/CRISP data, has disclosed useful information to explain this aspect. In fact, the negative magnetic flux decreases in the penumbral sector which firstly disappears on September~4 (the northern one, inside the box~A) only. This is accompanied by an almost constant positive magnetic flux. Conversely, for the remaining penumbral sectors, there are no significant changes in the penumbral field, although a similar quantity of positive magnetic flux is found. This provides a robust suggestion that the interaction between opposity polarity fields (type III MMFs) and the penumbral field has a key role during the decay process observed in this AR. On the other hand, benefiting from the extrapolation shown in Fig.~\ref{fig:estrapolazione}, we can argue that the different observed connectivities (red, white and green field lines in the extrapolation) may play a role in the onset of the penumbral decay, providing a possible explanation about why the penumbra disappears in sectors and not as a whole. Indeed, the different flux systems overlying the penumbra seem to be topologically separated, and this can be linked to the different instant of disconnection of the canopies leading to penumbral disappearance. 

The structure where small-scale opposite polarity features coexist, seen in particular on September~5 in the SST/CRISP observations, could be explained with the schematic illustration resulting from the numerical simulations presented by \citet{Kitiashvili2010}, where the sea-serpent configuration results as a consequence of solar overturning magnetoconvection in a highly inclined, strong magnetic field. In particular, the inclined magnetic field deforms the penumbral convective cells, the field lines are stretched and dragged down so that touching the surface form magnetic positive and negative polarities patches.

Finally, what role does the overlying canopy play in the decay of penumbra? The simulation work presented by \citet{Rempel2015} suggested how the strength of the mixed polarity field is based on the strength of the canopy field. Hence, this canopy field, being more prominent for spots than for pores, can have different effects on the size and nature of the region. In the present study, it is difficult to identify whether it initially dominates by triggering the onset of the decay in some part of the penumbra over others or it also plays a role by enhancing the mixed polarity fields leading to flux decay as suggested by the simulations.

The present work brings a valuable contribution to understand the penumbral decay process. However, it is clear that we have not been able to disentangle between the three mechanisms proposed to explain the process, although the role of MMFs and overlying canopies seem to have different spheres of action. Future observations by using space- and ground-based coordinated data, such as those from the PHI instrument \citep{Solanki2015,Solanki2020} on board of the Solar Orbiter mission \citep{Muller2013,Muller2020} and that from the existing (SST, GREGOR, and Goode Solar Telescope) and new 4-meter class telescopes (DKIST and EST) will be essential.

%\textcolor{red}{non colgo} \\
%\textcolor{blue}{da qui in poi rivedere}
%\salvo{ma non si vedono moti? l\'i in Romano2014 con IBIS qualcosa si vede...}
%\textcolor{blue}{FIN qui 04.03.2021 ore 10:15 \\ ultima parte da rivedere, solo abbozzi}
%\textcolor{teal}{commentare di pi\'u sta cosa delle estrapolazioni?}

%Our observations display also type III MMFs, i.e., with the opposite polarity with respect to the parent spot, moving along the penumbral field. \textbf{Therefore, we hypothesize that the decay process of the penumbra may be the result of the interaction of these MMFs and the penumbral field. This latter aspect is supported by the decreases of the signal in the Stokes V profiles. \color{red} Non possiamo dire che la variazione nell'intensità del segnale del profilo di Stokes mostri che il decadimento della penombra sia legato all'interazione delle MMF con il campo della penombra!!} \\
%\textcolor{blue}{questo sarebbe lo svolgimento dell'item (3)???}

%\textcolor{blue}{da qui in poi vedo solo sprazzi...}
%The plausible physical mechanism could be the submergence of the penumbral field due to the interaction of this latter and the type III MMFs, that are originated from the intersection of the submerged penumbral flux. \textbf{The main ingredient behind the decay process seems to be the interaction of the MMFs and the penumbral field. Indeed, when the MMFs are not detected the penumbral sector does not disappear. \color{red} Non si capisce cosa vuoi dire! In ogni caso quello che si dice nelle conclusioni deve essere supportato da risultati.}

\begin{acknowledgements}
The authors are grateful to the referee Dr. Horst Balthasar for the constructive comments.
This research received funding from the European Union’s Horizon 2020 Research and Innovation
531 program under grant agreements No 824135 (SOLARNET) and No 729500 (PRE-EST). This work was supported by the Italian MIUR-PRIN grant 2017 "Circumterrestrial environment: Impact of Sun-Earth Interaction" and by the Istituto NAzionale di Astrofisica (INAF). \\
SJ acknowledges support from the European Research Council under the European Union Horizon 2020 research and innovation program (grant agreement No. 682462) and from the Research Council of Norway through its Center of Excellence scheme (project No. 262622). \\
The authors wish to thank Dr. B. Ruiz Cobo for assisting us with the deconvolution of the HINODE data. The authors also thank Dr. J. M. Borrero for its helps in installing and preparing the VFSIV code. 
The Swedish 1-m Solar Telescope is operated on the island of La Palma by the Institute for Solar Physics of Stockholm University in the Spanish Observatorio del Roque de los Muchachos of the Instituto de Astrof{\'\i}sica de Canarias.
The Institute for Solar Physics is supported by a grant for research infrastructures of national importance from the Swedish Research Council (registration number 2017-00625).
This research is supported by the Research Council of Norway, project number 250810, and through its Centres of Excellence scheme, project number 262622.

\end{acknowledgements}


\end{document}